\documentclass[10pt,journal,compsoc]{IEEEtran}

\usepackage{ifpdf}

\ifpdf
  \pdfoutput=1\relax                   
  \usepackage{graphicx}                
  \DeclareGraphicsExtensions{.pdf,.png,.jpg,.jpeg} 
  \usepackage{graphicx}                
  \DeclareGraphicsExtensions{.eps}     

\graphicspath{{figures/}{pictures/}{images/}{./}} 

\usepackage{microtype}                 
\PassOptionsToPackage{warn}{textcomp}  
\usepackage{textcomp}                  
\usepackage{mathptmx}                  
\usepackage{times}                     
\usepackage{cite}                      
\usepackage{tabu}                      
\usepackage{booktabs}                  
\usepackage{amssymb}
\usepackage{enumitem}          
\usepackage{wrapfig}
\usepackage{graphicx}

\usepackage{soul}
\usepackage[dvipsnames]{xcolor}

\definecolor{darkpastelgreen}{rgb}{0.01, 0.75, 0.24}

\newif\ifproofread

\usepackage{xspace}
\newcommand{\etal}{\emph{et al.}\@\xspace}
\newcommand{\ie}{\emph{i.e.}\xspace}
\newcommand{\eg}{\emph{e.g.}\xspace}
\newcommand{\etc}{\emph{etcetera}\@\xspace}
\newcommand{\etals}{\mbox{\emph{et~al.}'s }}

\newcommand{\CDF}{\textsc{cdf}}
\newcommand{\CDFs}{\textsc{cdf}s}
\newcommand{\Barv}{\textit{Bar}}
\newcommand{\Pie}{\textit{Pie}}
\newcommand{\Bubble}{\textit{Bubble}}
\newcommand{\StackedBar}{\textit{Stacked Bar}}

\usepackage[utf8]{inputenc}
\usepackage{csquotes}

 \proofreadfalse  

\usepackage{amsmath}
\usepackage[english]{babel}

\usepackage{hyperref}
\hypersetup{unicode=true,
            pdftitle={The Risks of Ranking: Revisiting Graphical Perception to Model Individual Differences in Visualization Performance},
            pdfborder={0 0 0},
            breaklinks=true}
\usepackage{fancyvrb}

\DefineVerbatimEnvironment{Highlighting}{Verbatim}{commandchars=\\\{\}}

\usepackage{framed}
\definecolor{shadecolor}{RGB}{248,248,248}

\newcommand{\CommentTok}[1]{\textcolor[rgb]{0.56,0.35,0.01}{\textit{#1}}}

\newcommand{\DataTypeTok}[1]{\textcolor[rgb]{0.13,0.29,0.53}{#1}}

\newcommand{\KeywordTok}[1]{\textcolor[rgb]{0.13,0.29,0.53}{\textbf{#1}}}
\newcommand{\NormalTok}[1]{#1}
\newcommand{\OperatorTok}[1]{\textcolor[rgb]{0.81,0.36,0.00}{\textbf{#1}}}

\newcommand{\StringTok}[1]{\textcolor[rgb]{0.31,0.60,0.02}{#1}}

\newcommand{\codetilde}{\raisebox{.25ex}{\hbox{\texttildelow}}}

  \usepackage{cite}
\graphicspath{{../pdf/}{../jpeg/}}
\DeclareGraphicsExtensions{.pdf,.jpeg,.png}
  \usepackage{graphicx}
\begin{document}
%
\title {The Risks of Ranking: Revisiting Graphical Perception to Model Individual Differences in Visualization Performance}
%
%
%
%

\author{Russell Davis, Xiaoying Pu, Yiren Ding, Brian D. Hall, Karen Bonilla, Mi Feng, \\ Matthew Kay, and Lane Harrison

\IEEEcompsocitemizethanks{
\IEEEcompsocthanksitem Russell Davis, Yiren Ding, Karen Bonilla, Mi Feng and Lane Harrison are with Worcester Polytechnic Institute.\protect\\
E-mail: [rdavis, yding5, kbonilla, mfeng2, ltharrison]@wpi.edu.

\IEEEcompsocthanksitem Xiaoying Pu is with University of California, Merced.\protect\\
E-mail: xpu@ucmerced.edu.

\IEEEcompsocthanksitem Brian D. Hall is with University of Michigan.\protect\\
E-mail: briandh@umich.edu.

\IEEEcompsocthanksitem Matthew Kay is with Northwestern University.\protect\\
E-mail: mjskay@northwestern.edu.

}
}

%
%

\markboth{}%
{Davis \MakeLowercase{\textit{et al.}}: \thetitle}


\IEEEtitleabstractindextext{%
\begin{abstract}
Graphical perception studies typically measure visualization encoding effectiveness using the error of an ``average observer'', leading to canonical rankings of encodings for numerical attributes: \eg, position $>$ area $>$ angle $>$  volume. 
Yet different people may vary in their ability to read different visualization types, leading to variance in this ranking across individuals not captured by population-level metrics using ``average observer'' models. 
One way we can bridge this gap is by recasting classic visual perception tasks as tools for assessing individual performance, in addition to overall visualization performance.
In this paper we replicate and extend Cleveland and McGill’s graphical comparison experiment using Bayesian multilevel regression, using these models to explore individual differences in visualization skill from multiple perspectives. 
The results from experiments and modeling indicate that some people show patterns of accuracy that credibly deviate from the canonical rankings of visualization effectiveness.
We discuss implications of these findings, such as a need for new ways to communicate visualization effectiveness to designers, how patterns in individuals’ responses may show systematic biases and strategies in visualization judgment, and how recasting classic visual perception tasks as tools for assessing individual performance may offer new ways to quantify aspects of visualization literacy.
Experiment data, source code, and analysis scripts are available at the following repository:
\url{https://osf.io/8ub7t/?view\_only=9be4798797404a4397be3c6fc2a68cc0}.

\end{abstract}

}

\maketitle
\IEEEdisplaynontitleabstractindextext

%
\IEEEpeerreviewmaketitle

\IEEEraisesectionheading{\section{Introduction}\label{sec:introduction}}

\IEEEPARstart{V}{isualizations} continue to be created for, and read by, a broad and diverse audience.
Advances in visualization authoring tools alongside the rise of social media
have contributed to a greater saturation of visualizations in peoples' daily lives.
One challenge posed by this increased exposure is that people may vary in their ability to perform fundamental visualization tasks, such as estimating and comparing values, judging correlations, or identifying trends and outliers.
While there are some situations in which people might change a visualization to suit their needs (\eg shared spreadsheets), there are numerous everyday scenarios such as digital journalism, television, newspapers/magazines, and public settings where the decisions designers make about visual encodings cannot be changed.
Complicating the problem is the observation that we know little about the extent to which and in what ways people can vary in visualization performance, as many studies focus on the performance of the ``average observer'' rather than on the variance in participants themselves.

Graphical perception studies are one of the primary means through which visualization research has developed a better understanding of how people perform fundamental tasks with visualizations, stretching back to the classic work of Cleveland and McGill~\cite{cleveland1984graphical}.
Such studies have yielded several longstanding results, such as canonical rankings of visualization effectiveness~\cite{cleveland1984graphical}, which form the basis of visualization guidelines \cite{munzner2014visualization, berinato2016good} and recommendation systems \cite{mackinlay1986automating, moritz2018formalizing}. Other graphical perception studies have validated the use of crowdsourcing to evaluate visualizations~\cite{heer2010crowdsourcing}, and explored how social information can influence visualization judgments~\cite{hullman2011impact}.
Yet because the majority of graphical perception studies assess quantitative performance based on the ``average observer'', it is difficult or impossible to gain an understanding of how people vary in performance at the individual level. 
Moving beyond an exclusive focus on the ``average observer'' could allow us, for example, to assess how individuals differ from the broader population, or to understand how consistently canonical rankings of visualization effectiveness hold across a wide range of people.

As Cleveland and McGill acknowledge in their original paper on graphical perception~\cite{cleveland1984graphical}, modeling the variance in individual performance across a population is a substantial undertaking:
\begin{displayquote}
Because each subject judged all of the experimental units in an experiment, the judgments of one unit are correlated with those of another, and modeling this correlation would have been a substantial chore.
\end{displayquote}
Equipped with recent advances in Bayesian methods (such as modern Markov chain Monte Carlo samplers \cite{carpenter2017stan}) that have made it easier to fit complex hierarchical models, we aim to undertake that chore.
These techniques make it possible to fit models to individual-level observations---estimating individual-level parameters for the mean and variance of a person's observations and the correlation between those parameters---while simultaneously estimating population-level means and variance.
Such models are sometimes called mixed effects location-scale models (MELSM) \cite{Hedeker@2008melsm}, or multilevel distributional regression \cite{burkner2017advanced}.

\begin{figure*}[t]
    \centering
    \includegraphics[width=1.0\linewidth]{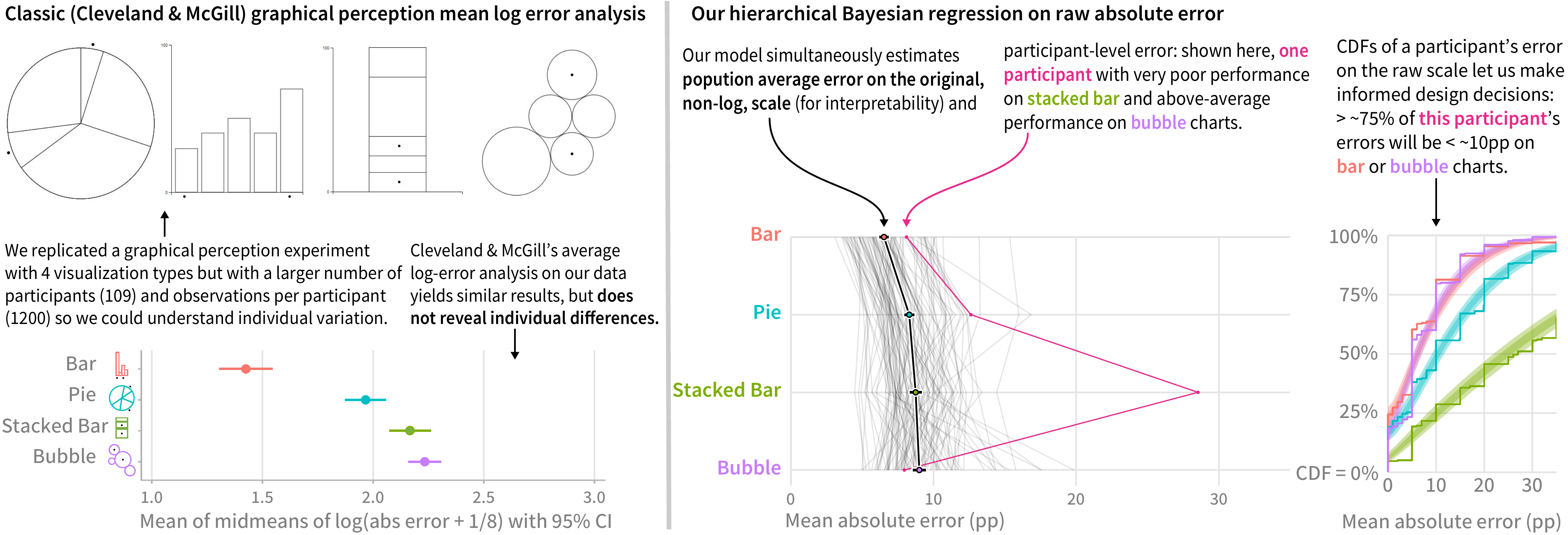}
    \caption{A comparison of a Cleveland and McGill-style average log error analysis on our data with our hierarchical analysis.}
    \label{fig:teaser}
\end{figure*}

Recently, researchers have applied MELSM techniques to understand variance in individuals' performance on some specific visualization types, such as uncertainty visualizations \cite{fernandes2018uncertainty} or visualizations of correlation \cite{kay@2016beyond}. 
A recent study from McColeman \etal targets ratio perception tasks that underlie the comparison task in Cleveland and McGill's study, finding deviations in performance from traditional ranks of visual channels \cite{mccoleman2021rethinking}.
However, the fundamental comparison tasks from Cleveland and McGill \cite{cleveland1984graphical}, which underlie many replication efforts in visualization as well as visualization design recommendations, have not been examined from the perspective of individual variance. 
This raises a question: might individual people substantially vary in performance on the same tasks that form the basis of widespread guidance in visualization design? 

In this paper, we investigate individual differences in graphical perception through a crowdsourced experiment and Bayesian modeling.
We replicate and extend the comparison task from Cleveland and McGill's study, 
making necessary adjustments to facilitate person-to-person comparison, such as consistent stimuli sets and repeated trials.
Using a total of 130,800 judgments from 109 crowdsourced participants, we progress through a series of models, beginning with mean log error approaches from prior studies and ending with a model of absolute error in individual observations with participant-level effects for each visualization type and correlations between them. 
Our contributions include:

\begin{itemize}
\item \textbf{Evidence of substantial differences in peoples' graphical perception performance.} Results from the proposed hierarchical models suggest that people can vary considerably compared to the ``average participant''. On certain chart types, some perform consistently better (up to 5pp), while others perform worse (up to 20pp), see Figure \ref{fig:individual-summary}.
This pattern holds across the tested population. Results show that expected differences in performance across pie, bubble, and stacked-bar charts (1-1.5pp) is smaller than the expected differences between people (1.5-3pp), see Figure \ref{fig:variance-between-people-and-charts}.

\item \textbf{Positive correlation in performance across visualization types.} Fulfilling the ``chore'' of correlation modeling as described by Cleveland and McGill \cite{cleveland1984graphical}, we estimate correlations in individuals' performance between chart types. We find that performance is generally positively correlated across all chart types ($r\approx0.5-0.7$), 
though this correlation is weak for some pairs, \eg \StackedBar{} and \Bubble{} ($r\approx0.3$), see Figure \ref{fig:individual-correlation}.

\item \textbf{What is ranked best for the average participant may be not ranked best for a substantial portion of people.} While \Barv{} charts elicit the best performance on average, results show that around 20-25\% of people consistently perform better with another chart type. Less than 40\% of people are expected to share the ``canonical'' ranking of \Barv{} best and \Pie{} as second-best, with some instead performing better with \StackedBar{} or \Bubble{}, see Figure \ref{fig:ranking-uncertainty}.
\textbf{This calls into question whether we should be using rankings at all to derive design guidance}. Instead, we might encourage the use of effect sizes and their uncertainty to make judgments about practical differences in encodings for a given visualization design context.

\end{itemize}

The variation in performance between individuals across different chart types may imply that the role of psychophysical explanations for differences in error rates has been overstated. In addition to the prior points, many participants perform very similarly with chart types that are supposed to possess very different psychophysical properties, and the effects of variance across individuals can dwarf the mean effects of chart type. We explore how these findings suggest that individual-level factors might be more important to attend to than suggested by prior research, yet the precise nature and origin of these factors are unknown.

Considering visualization design, we discuss how graphical perception research can be made more accessible by shifting reporting more towards constructs such as the variance in people's performance and the use of raw error measures (instead of log error). 
There is also a need to improve how we measure and communicate visualization effectiveness to designers beyond the ``average'' participant.
We take a step towards this goal by exploring model-driven probabilistic rank representations (see Figure \ref{fig:ranking-uncertainty}).
Finally, we discuss how these findings and the modeling approaches that drive them may provide needed support for the expanding visualization literacy efforts in our community.

\section{Background}

Graphical perception studies evaluating how people perform basic visualization tasks are a common refrain in visualization research.
For example, these studies have been used to investigate particular visualization techniques, like treemaps \cite{kong2010perceptual}, variants of bar charts \cite{srinivasan2018s}, or aspects of peoples' behavior with visualization, like social bias \cite{hullman2011impact}.
We draw on prior graphical perception studies, as well as work in visualization and human-computer interaction that has moved beyond population level-analyses.

\subsection{Graphical Perception Studies in Visualization}

In a study design common in graphical perception research, experimenters vary \emph{visualization types} (\eg, \Barv{}, \Pie{}, and \Bubble{} charts) to measure the effect of different ways of encoding data on participants' error in reading that data~\cite{cleveland1984graphical,heer2010crowdsourcing,heer2009sizing}.
Cleveland and McGill's seminal graphical perception study examined several ``elementary \emph{perceptual tasks}'' across different visualization types~\cite{cleveland1984graphical}.
Using 95\% bootstrapped confidence intervals on trimmed means of log error, they derived design recommendations based on participants' average performance across different visualization types. 

Heer \etal~\cite{heer2010crowdsourcing} and others (\eg \cite{hullman2011impact}) have replicated parts of Cleveland and McGill's original study for various purposes, including validating that online platforms such as Mechanical Turk (MTurk) are a viable testbed for conducting graphical perception studies~\cite{heer2010crowdsourcing}. 
Cleveland and McGill's results have also been cited in the development of tools for automatically creating effective visualizations~\cite{mackinlay1986automating}, thereby having influence on visualization design practice.

Other studies using similar visualization types but different tasks contest the visual encoding rankings from Cleveland and McGill's original study. 
Hollands and Spence, for example, demonstrate that the original ranking is not necessarily suitable in explaining the average participants' performance when it comes to ``discrimination'' tasks, \ie discerning the larger of two quantities \cite{hollands2001discrimination}. 
Similarly, Yuan \etal found that the ranking also does not apply in tasks requiring a comparison of multiple values, \eg averages across various data points \cite{yuan2019perceptual}.
Chung \etal investigate and rank the perceptual orderability of visual channels such as hue, size, texture, \etc. \cite{chung2016ordered}.

As these prior studies rank visualizations based on means (representing the ``average observer''), other informative measures, such as variance---how people may differ across visualization types, and how consistently individuals make the same judgment---are less apparent. 
One aim of our work is to advocate for more analysis of individual-level phenomena in graphical perception, as opposed to only drawing conclusions about the average of a population. 
As argued by Ziemkiewicz and Kosara, a better understanding of the possible disparities and differences between how people perform with visualizations could lead to more thorough incorporation of individual differences into the design of systems used to create visualizations \cite{ziemkiewicz2009preconceptions}. 

\begin{figure*}[ht]
    \centering
    \includegraphics[width=\linewidth]{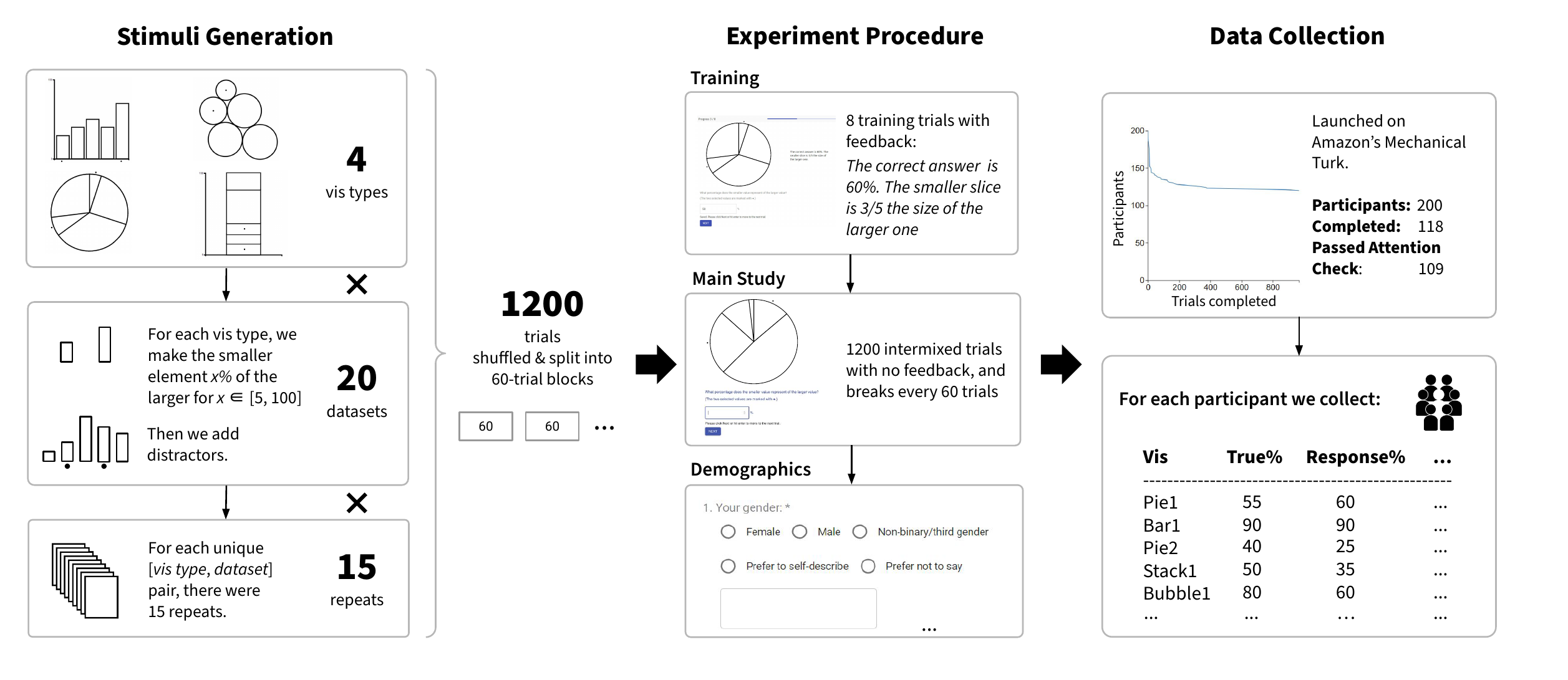}
    \caption{Experiment Overview: To facilitate person to person comparison in a graphical perception experiment, we make necessary changes to the stimuli and data generation processes, illustrated here. The experiment procedure and data collection details are also shown.}
    \label{fig:experimentflow}
\end{figure*}

\subsection{Individual Differences in Visualization}

The term ``individual differences'' in visualization typically refers to differentiating visualization performance on the basis of factors like personality traits, scores on cognitive ability tests, cognitive states, \etc.
However, individual differences might also refer to examining variance between participants themselves. 
Binning participants by personality factors or spatial ability scores, for example, will not necessarily reveal performance differences between individuals when groups are the focus of the analysis.
In the present work, we focus on measuring between-participant variance and making performance comparisons at the individual level, while contributing statistical models that realize this goal.

\subsubsection{Individual Differences as Subgroup Analyses}

Several studies have focused on ``individual differences'' as ``traits or stable tendencies to respond to certain classes of stimuli or situations in predictable ways'' \cite{dillon1996user, liu2020survey}. 
For example, Ziemkiewicz and Kosara explore personality factors such as extraversion and openness, finding impacts on task performance measures such as speed and accuracy \cite{ziemkiewicz2009preconceptions}.
Further studies from Green and Fisher, Ziemkiewicz \etal, and others explore the impact of personality traits in a variety of visualization contexts \cite{green2010towards, ziemkiewicz2012visualization}.
Similar effects are found for spatial ability by Micallef \etal and Ottley \etal in Bayesian reasoning tasks with visualizations \cite{micallef2012assessing, ottley2015improving}.
Liu \etal provides a comprehensive survey of these and more individual differences-focused studies in visualization \cite{liu2020survey}.

\subsubsection{Average vs. Individual Level Analyses in Visualization}

Mean performance, such as the perceptual error of an ``average observer'', is a concise measure that can be extrapolated over a population and can easily be transformed into a guideline, \ie ``just pick the visualization type with lowest average error!'' 
Yet if we only rely on mean performance to rank the effectiveness of visualizations for a given task, there is a possibility that trends for the average observer may deviate at the individual level.
Fortunately, recent studies in visualization have begun to use hierarchical modeling and Bayesian techniques to better account for individual variance in population level measurements, laying a foundation for thinking beyond the average.

Harrison \etal modeled peoples' perception of correlation differences in several bivariate visualization types using Weber's law, providing a means to quantitatively evaluate (and thus rank) the effectiveness of each tested type \cite{harrison@2014ranking}. 
Kay \etal built on Harrison \etals work by applying Bayesian modeling to incorporate variation between individuals using random intercepts \cite{kay@2016beyond}.
Kay \etal proposed a new ranking of visualizations for correlation discrimination that incorporated model-derived uncertainty, and in evaluating this issue at the individual level, they concluded that canonical rankings in visualization may unintentionally ``overstate the strength of the evidence'' they are based on. 

Beecham \etal quantified peoples' confidence in drawing conclusions from geospatial data visualizations---using a similar methodology as Harrison \etals~\cite{harrison@2014ranking}'s JND experiments \cite{beecham2016map}. 
Following Kay \etal~\cite{kay@2016beyond}, Beecham \etal also included a random intercept to account for differences in performance between participants, improving model fit in one of their conditions. 
Such hierarchical models have also been applied to understanding variance in peoples' performance on different types of uncertainty visualizations~\cite{fernandes2018uncertainty,kale2020visual}.
Additionally, Lu \etal have extended these modeling efforts to explore differences in discriminability for bar, bubble, and pie charts \cite{lu2021modeling}.

\subsection{Perception Studies}
Studies in perceptual psychology and vision science have also introduced modeling results involving graphical elements common in visualizations.
Early work in psychophysics, such as Fechner's introduction of Weber's law \cite{adler1966elements}, was cited as partial inspiration for Cleveland and McGill's graphical perception experiments, and used directly in recent work in the visualization field for modeling the perception of correlation in scatterplots (\eg \cite{rensink2010perception, harrison@2014ranking, kay@2016beyond}). 
Other psychophysics research has dealt with ratio estimation problems directly.
Stevens reviews several methodologies and experiment paradigms in psychophysics, including production and estimation techniques for ratios and magnitudes \cite{stevens1958problems}.
Ekman investigates ratio estimation procedures and discusses model fitting procedures to account for potential biases in participant responses \cite{ekman1958two}. 
Ekman \etal also investigate interindividual differences in perceptual ratio estimation in stimuli such as weight, brightness, and area, with results suggesting that participants can vary systematically at the perceptual level \cite{ekman1968interindividual}.
Baird reports a bias towards multiples (\eg 1, 5, 10) in free-response ratio estimation tasks \cite{baird1970relative}, an effect which has been reported in Talbot \etals partial replication of Cleveland and McGill's graphical perception study \cite{talbot2014four}.

A common goal throughout these studies is investigating peoples' responses to various visual stimuli. We draw on these experiment methodologies and modeling considerations for the current experiment. 
In contrast to prior perception-focused experiments, we intentionally adopt Cleveland and McGill's task because of its perceptual and cognitive dimensions.
A perception focused study, for example, might use the method of quadruples where two pairs of charts are shown and the participant is asked to identify which shows the higher ratio \cite{mlds}.
However, we note that the original estimation task is closer to how people use charts in daily practice, for example comparing slices of an individual pie chart in the boardroom. 
These cognitive dimensions of the task are a potential source for differences in chart reading ability, which we aim to investigate.

Taking these insights and motivations for studying between-person variance, adopting more sophisticated statistical methods such as Bayesian hierarchical regression, and returning to the classic graphical perception tasks of Cleveland and McGill, we aim to pursue a more principled, flexible, and robust approach to asking how much individuals vary in their ability to perform basic graphical perception tasks with visualizations.

\section{Methodology}

The primary motivation of this study is to explore how people might vary in their ability to perform basic visualization tasks.
To do so, we adapt and extend the graphical comparison experiment from Cleveland and McGill~\cite{cleveland1984graphical}. We then use a hierarchical Bayesian model to examine variance in performance both between individuals and between visualization types, as well as correlations between participants' performance across visualization types.
This graphical comparison experiment has been previously replicated in a range of studies targeting different visualization types and participant scenarios (\eg~\cite{hullman2011impact, harrison2013influencing, talbot2014four, heer2010crowdsourcing, kong2010perceptual}). 
As such, it also serves as a touch point for broader discussion on modeling approaches in visualization.

In addition, close examination of the methodologies in these prior experiments reveals that they are not directly suitable for person-to-person performance comparison. 
For example, some participants might see tasks from only one chart type, requiring between-subjects comparison.
To address these issues and to facilitate our goal of comparing person-to-person performance, we highlight several minor but necessary changes to the experiment protocol (see Figure \ref{fig:experimentflow} for an overview).

\subsection{Experiment Stimuli and Data Generation}

The majority of the experiment protocol is adapted from Cleveland and McGill's original design~\cite{cleveland1984graphical} and Heer and Bostock's crowdsourced replication of it~\cite{heer2010crowdsourcing}. 
Specifically, participants still see plain, black and white visualizations of five data points--- two marked for comparison--- and give a numerical answer to ``what percentage is the smaller of the larger?''.
One difference in our experiment are the visualization types tested, which needed to be chosen to facilitate performance comparisons between participants while not excessively expanding the experiment length.
In Cleveland and McGill's comparison experiment, five variations of chart types were tested, derived from \Barv{} and \StackedBar{} charts.
Heer and Bostock's replication extended this count to 9 types, adding \Pie{} charts, \Bubble{} charts, stand-alone rectangular areas, and treemaps.

To select charts that would allow us to compare performance between people, we analyzed the results reported in the studies, with two selection criteria in mind.
The first goal was to choose a visualization that could serve as a baseline across all participants.
The \Barv{} chart, given its consistently lower error relative to other chart types in these studies and other replications, met this criterion.
The second goal was to choose a small set of visualizations that were similar in average performance, to raise the possibility of identifying people who perform consistently better or worse on them in contrast to the canonical ranking (shown in Figure \ref{fig:teaser}). 
Given their similar performance in prior studies, \StackedBar{}, \Pie{}, and \Bubble{} charts met this criterion.
Other factors were weighed in the decision-making process, such as evidence from Kosara \etal~\cite{kosara2019evidence} that people may use different strategies when judging \Pie{} charts, which could lead to systematic performance differences, and Kong \etal's study showing that orientation issues lead to difficulties in interpreting treemaps~\cite{kong2010perceptual} (we exclude treemaps for this reason). 

\begin{figure}
    \centering
    \includegraphics[width=0.6\linewidth]{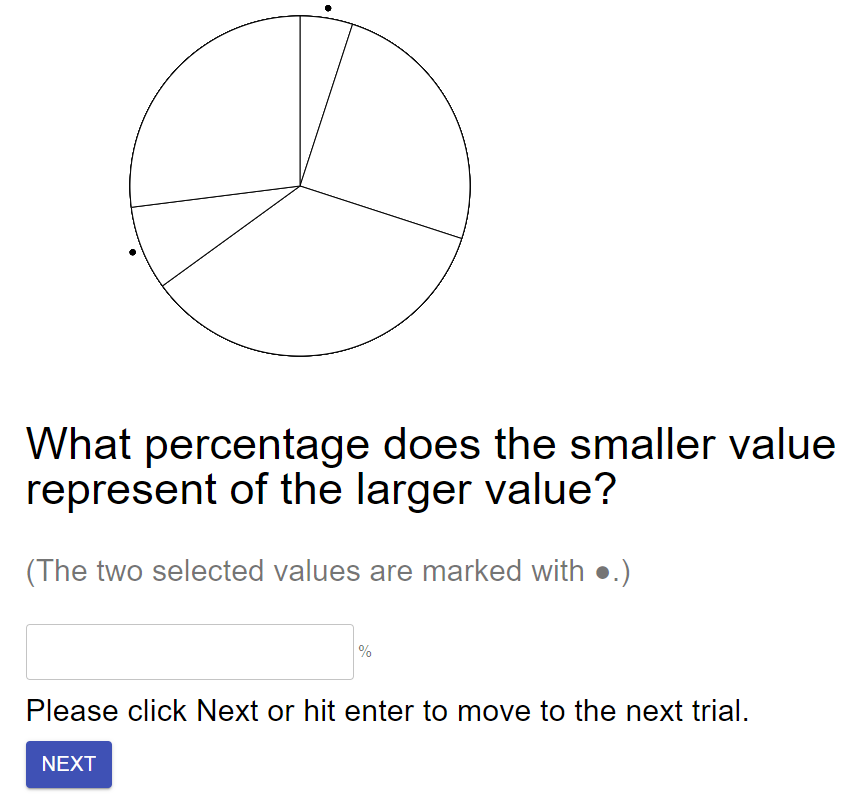}
    \caption{An example task from the experiment: a \Pie{} chart with two slices marked for comparison with $\bullet$.}
    \label{fig:task}
\end{figure}

The data generation process follows prior studies \cite{cleveland1984graphical, heer2010crowdsourcing}, with modifications to accommodate repeated trials and consistent stimuli across participants for participant-to-participant comparison. 
Five values are generated in each trial, including 1 smaller comparison value ($S$), 1 larger comparison value ($L$), and 3 distractor values. 
To evenly cover the domain of possible answers, the $proportion$ of the smaller value to the larger value is separated by 5\%, ranging from 5\% to 95\% (19 values), plus a 99\% for a total of 20 different proportions. 
The 5\% differences in stimuli is informed by prior studies, in particular Talbot \etal, who find that participants typically give answers ending in 5 or 0 \cite{talbot2014four}.
For the \Barv{}, \Bubble{}, and \StackedBar{} charts, the 3 distractor values are randomly generated within a normalized 0 to 1 range. 
The \Pie{} chart, given its part-to-whole arrangement, is handled differently: all 5 values are constrained to sum to 1 to represent the whole of the chart. Figure \ref{fig:task} shows an experiment task with the pie chart stimuli.
In total, for each \emph{visualization-dataset} pair ($4x20$), there were 15 \emph{repetitions}.
Thus each participant answered 1200 trials in total (4 \emph{visualizations} $\times$ 20 \emph{datasets} $\times$ 15 \emph{repetitions}).
After the set of trials was created, all were shuffled to appear in random order.

\subsection{Experiment Procedure} \label{sec:experiment-procedure}

The experiment procedure included three phases: \emph{Training}, \emph{Trials}, and \emph{Demographics}. 

\textbf{Training}:
Participants were given eight practice trials, covering all four visualization types.
After answering, participants were shown feedback text with the correct answer.
For example, in a \Barv{} chart practice trial, participants might see: \emph{The correct answer is 42\%. The smaller bar is just over 2/5 the size of the larger one.}
Training trials included a mixture of rounded and non-rounded answers, to emphasize to the participant that the answers could feasibly take on any value between 1 and 100.

\textbf{Trials}:
The trials phase began with a paragraph that reminded the participant that their answers were being recorded.
As in prior experiments~\cite{heer2010crowdsourcing}, participants were encouraged to make a ``quick visual judgment'' and to avoid physically measuring the stimuli to get exact answers.
Breaks were given after every 60 trials (20 breaks in total). 
The participants were encouraged to take a break as long as needed before resuming.

\textbf{Demographics}:
After participants finished all training and trials, they were asked to provide basic demographic information, including reported gender, age, country of origin, and highest degree obtained.
Participants were also asked to self-rate (on a 1-7 scale) their experience with visualization and statistics.

\subsection{Resulting Participants \& Exclusion Criteria}
Participants were recruited in an IRB-approved \footnote{WPI IRB Review Board \#16-176M4} study on Amazon's Mechanical Turk, which has been shown to be a reliable testbed for graphical perception experiments \cite{heer2010crowdsourcing}. 
200 participants started the experiment, and 118 of them completed all 1200 trials. 
A survival chart is shown in the data collection section of Figure \ref{fig:experimentflow}, indicating that the experiment had a 59\% completion rate in total. 
Among the participants who did not finish all trials, 57.3\% dropped before 10 trials and 80.5\% dropped before 100 trials.
Each participant received the same set of trials, making this a within-subjects design. Each participant that completed all trials was compensated \$22. 
The average completion time was 2 hours 40 minutes, for an hourly rate \$8.3/h, exceeding U.S. minimum wage.

Conditions where the true proportion was 100\% (both elements were the same size) and 5\% are additionally used as attention checks.
While a range of errors can be expected, repeated large errors on these trials likely represent a participant not paying attention, not trying to be accurate, or misunderstanding the experimental instructions.
We only excluded participants where extreme errors of greater than 50pp occurred on more than 25\% of trials in either of the two conditions, which equates to 16 or more such occurrences out of 60 trials per condition. 
A total of 9 participants met these criteria, and were thus excluded from all further analyses, leaving 109 participants in all. Data with and without such exclusions are available in supplemental material.

\section{Analysis Approaches}

We will use two approaches to analyzing the data: a partial replication of Cleveland \& McGill's analysis~\cite{cleveland1984graphical} which focuses on means, then a Bayesian regression model that allows us to also examine within- and between-participant variance. 
We will use a \textit{model expansion} to develop a model that describes, as well as we are able, the phenomena in question, rather than proposing and testing specific hypotheses about the data~\cite{gelman2020bayesian}.
Before describing the results (Section \ref{sec:results}), we will first describe both approaches in detail.

\subsection{Replicating Cleveland \& McGill}

We replicate Cleveland \& McGill's analysis
\cite{cleveland1984graphical} to put our study in context with their
seminal findings. In addition, since Heer \& Bostock closely replicates
Cleveland \& McGill, we will compare findings from both
papers in Section \ref{result-replicate}. For the replication analysis, we use the same formula for log error as Cleveland \& McGill:
\(\log_2( \left| \text{judged percent} - \text{true percent} \right| + {1}/{8})\).
At a high level, the analysis uses bootstrapping to calculate means and confidence intervals of midmeans\footnote{The \emph{midmean} or \emph{interquartile mean} is the mean of the central 50\% of the data (the data between the first and third quartiles).} of log errors (see Section 4.3 and 4.4 in the original paper
\cite{cleveland1984graphical}). First, we create 1000 bootstrapped
samples. Each sample has 109 participants
resampled from our data with replacement (109 $\times$ 1200 = 130800 observation). For each sample:

\begin{enumerate}[noitemsep]
\def\labelenumi{\arabic{enumi}.}
\item
  Calculate log error = $\textrm{log}\left(|\textrm{error}|+\frac{1}{8}\right)$ for each observation.
\item
  In Cleveland \& McGill~\cite{cleveland1984graphical} and Heer \& Bostock~\cite{heer2010crowdsourcing}, each participant completes
  one trial per condition combination (true proportion and visualization
  type), while we ask each participant to complete 15 repeated trials
  per combination. Thus, to be most comparable to previous approaches,
  we compute the mean response from the 15 repeated trials for each combination of 
  true proportion, visualization, and participant.
\item Take mid-means within each combination of visualization type and true proportion, yielding 80 mid-means (4 visualizations $\times$ 20 true proportions).
\item
  Group by visualization type and take the mean in each group, yielding four means of midmeans, one for each visualization type.
\end{enumerate}
The result of the above procedure is a bootstrapped sampling distribution of 1000
means of midmeans of log errors for each visualization type, from which we can
calculate 95\% confidence intervals.

\subsection{Building a more complete model of errors} \label{sec:model-explanation}

To build up to a more complete model---one which describes absolute errors at the visualization and participant level, allowing individuals' abilities (and consistencies) to vary between each other and across visualization types---we will first start with a simplified model. For interpretability, we will also aim to have a model that describes errors on the original percentage response scale, not on a log scale as in Cleveland and McGill~\cite{cleveland1984graphical}. To develop our final model, we followed a \emph{model expansion} approach~\cite{gelman2020bayesian}, gradually adding more complexity to the model to describe the assumed data generation process in more detail. Throughout this process, we assessed model fit and quality using posterior predictive checks~\cite{gabry2019visualization} to understand in what ways a given model failed to describe the data generation process. We then used these checks to determine in what direction to expand the model until it was able to adequately describe the data. (In addition to the walkthrough here, we provide annotated source files for replicating this analysis and experiment in the supplemental material.)

We will begin with a model of mean absolute errors, assuming that errors have been averaged within \emph{participant} $\times$ \emph{vis}. In other words, for each participant, we calculate their mean absolute error on each visualization. Such a model allows us to look at mean absolute error at the visualization level (much like the Cleveland and McGill analysis), but does not permit us to analyze individual-level performance, as this is averaged out in advance. This is a common approach in the visualization literature (\eg \cite{heer2010crowdsourcing, hullman2011impact, talbot2014four, kong2010perceptual, harrison2013influencing}).

To build a Bayesian model, we also need a \emph{likelihood}. The likelihood is a distribution that we assume observations to be drawn from, conditional on the predictors in the model. Each observation in this model is the mean absolute error for one participant on a particular visualization condition. Traditional linear regression, for example, typically assumes a Normal likelihood:
\newcommand{\mae}{\textrm{mean\_abs\_error}}
\newcommand{\error}{\textrm{abs\_error}}
\newcommand{\Normal}{\textrm{Normal}}
\newcommand{\HalfNormal}{\textrm{half-Normal}}
\newcommand{\Student}{\textrm{Student\_t}}
\newcommand{\vis}{\textrm{vis}}
\newcommand{\ZIB}{\textrm{ZeroInflatedBeta}}
\newcommand{\Beta}{\textrm{Beta}}
\newcommand{\Bernoulli}{\textrm{Bernoulli}}
\newcommand{\logit}{\textrm{logit}}
\newcommand{\newcol}[1]{\begingroup\color{red}#1\endgroup}

\newcommand{\participant}{\textrm{participant}}

\begin{align*}
V = 4 &: \textrm{number of visualization types}\\
P = 109 &: \textrm{number of participants}\\
i \in \{1 \dots VP\}&: \textrm{index of observations}\\
&~~~ \textrm{(mean errors within participant} \times \textrm{vis)}\\
\vis[i] \in \{1 \dots V\} &: \textrm{visualization associated with the \textit{i}th observation}
\end{align*}
\begin{align*}
\mae[i] &\sim \Normal(\mu[i], \sigma) &&\textit{likelihood}\\
\mu[i] &= \beta[\vis[i]] &&\textit{mean submodel}
\end{align*}

This model says that each observation, $\mae[i]$, is Normally distributed with a mean of $\mu[i]$ and standard deviation of $\sigma$. Given a specific visualization type $v \in \{1 \dots V\}$, $\beta[v]$ is the average mean absolute error for that visualization type. Thus, $\beta[\vis[i]]$ is the average mean absolute error for the visualization type associated with observation $i$ in the dataset.

This model fails to capture individual-level differences: e.g., that some participants might be better or worse on some visualization types, or even systematically better or worse on these tasks in general. The Normal likelihood also fails to capture key constraints of the data generating process: e.g., we know that absolute error on a percentage response scale must be between $0\%$ and $100\%$ (i.e., 0 and 1). 

Let's tackle the latter problem first: adjusting the likelihood. Instead of the Normal likelihood, we'll adopt a \emph{zero-inflated Beta} distribution as the assumed distribution of errors, described by a \emph{mean} parameter ($\mu$), \emph{precision} parameter ($\phi$; also called the \emph{sample size} parameter, this gets larger as variance gets smaller), and a \emph{zero probability} parameter ($\pi$). The Beta distribution is defined on $(0, 1)$, and so is commonly used to model bounded data~\cite{smithson2006better,burkner2017brms}. Unfortunately, the Beta distribution does not allow zeros, yet our data contains zeros at the individual observation level (when a participant gets a response exactly correct). The zero-inflated Beta distribution is a modified Beta distribution that allows zeros by modeling the probability of a zero being present as a separate process, as follows:
\begin{align*}
    y &\sim \ZIB(\mu, \phi, \pi)\\\\
    \implies y &= \begin{cases}
            0 & \text{if $z$ = 1}\\
            y^{*} & \text{if $z$ = 0}
        \end{cases}\\
    y^{*} &\sim \Beta\left(\mu\phi, (1 - \mu)\phi\right)\\
    z &\sim \Bernoulli(\pi)
\end{align*}
Like the observations in a Beta distribution, its mean parameter ($\mu$) must also be between 0 and 1. Thus, to adjust our model to use the zero-inflated Beta distribution, we must ensure that $\mu$ is bounded between 0 and 1. 
We can do this by using a \emph{link function} that takes mean absolute errors in $(0,1)$ and translates them onto $(-\infty,+\infty)$; thus the inverse of this link function ensures $\mu$ is in $(0,1)$. The logit function does this (changes from the previous specification are highlighted in red):
\begin{align*}
    \mae[i] &\sim \newcol{\ZIB}(\mu[i], \phi, \pi) &&\textit{likelihood}\\
    \newcol{\logit(}\mu[i]\newcol{)} &= \beta[\vis[i]] &&\textit{mean submodel}
\end{align*}

To model individual errors directly, one might naively think to simply fit the above model to participant-level errors instead of mean errors. 
However, doing so would result in artificially inflating the number of samples in the model, leading to overconfident estimates (\emph{pseudoreplication}~\cite{hurlbert1984pseudoreplication}).
To address this problem, we must ensure that the model accounts for differences in individuals' performance. We can do this using \emph{random effects} by adding an offset $U[\vis[i],\participant[i]]$ that describes how the participant associated with observation $i$ deviates from the average for the visualization associated with that same observation.\footnote{In the terminology of \emph{random intercepts} and \emph{slopes}, this offset combines a per-participant random intercept with a random slope for each visualization conditional on participant.} 
Now we can model individual errors directly, without averaging first:

\begin{footnotesize}

\begin{align*}
V = 4 &: \textrm{number of visualization types}\\
P = 109 &: \textrm{number of participants}\\
\newcol{K = 300} &\newcol{: \textrm{number of repetitions}}\\
i \in \{1 \dots VP\newcol{K}\}&: \textrm{index of observations}\\
&~~~ \textrm{(\newcol{\st{mean} trial-level} errors)}\\
\vis[i] \in \{1 \dots V\} &: \textrm{visualization associated with observation \textit{i}}\\
\newcol{\participant[i] \in \{1 \dots P\}} &\newcol{: \textrm{participant associated with observation \textit{i}}}
\end{align*}
\begin{align*}
    \newcol{\textrm{\st{mean\_}}}\error[i] &\sim \ZIB(\mu[i], \phi, \pi) &&\textit{likelihood}\\
    \logit(\mu[i]) &= \beta[\vis[i]] \newcol{+ U[\vis[i], \participant[i]]} &&\textit{mean submodel}
\end{align*}
\end{footnotesize}

To add a random offset $U[v,p]$ for a particular visualization $v \in \{1 \dots V\}$ and participant $p \in \{1 \dots P\}$, we must also estimate the correlation between the random offsets for different visualization types, which allows us to understand (for example) if someone who performs better on one visualization type also tends to perform better or worse on another visualization type. Because the logit link function has transformed the means onto a latent scale between $(-\infty,+\infty)$, this can be done using a multivariate normal distribution, where the associations between random effects are captured by the covariance matrix $\Sigma$:
\begin{footnotesize}

\begin{align*}
    \newcol{\begin{bmatrix} U[1,p]\\ \vdots\\ U[V,p] \end{bmatrix}}
        &\newcol{\sim \Normal\left(
            \begin{bmatrix} 0\\ \vdots\\ 0 \end{bmatrix},
            \Sigma
        \right)}
        &&\newcol{\forall p \in \{1 \dots P\}}
        &\newcol{\begin{array}{l}\textit{correlated}\\ \textit{random offsets}\end{array}}
\end{align*}
\end{footnotesize}

While the mean ($\mu[i]$) is allowed to vary by visualization and participant in the above model, neither the precision ($\phi$) nor the probability of a zero ($\pi$) does. This is a strong assumption; relaxing it would allow the model to capture the fact that some people may be more or less \emph{consistent} in the size of errors they make. With respect to variance or precision parameters, relaxing this assumption is sometimes called accounting for \emph{heteroskedasticity}, which simply means that variance in some conditions (or for some people) may be different. We will add submodels for both $\phi$ and $\pi$ that echo the existing submodel for $\mu[i]$, with link functions that transform the latent scale onto the appropriate bounds for each parameter: $(0,+\infty)$ for $\phi$ (hence log) and $(0,1)$ for $\pi$ (hence logit):

\begin{footnotesize}

\begin{align*}
    \error[i] &\sim \ZIB(\mu[i], \phi\newcol{[i]}, \pi\newcol{[i]}) &&\textit{likelihood}\\
    \logit(\mu[i]) &= \beta_{\newcol{\mu}}[\vis[i]] + U_{\newcol{\mu}}[\vis[i], \participant[i]] &&\textit{mean submodel}\\
    \newcol{\log\left(\phi[i]\right)} &\newcol{= \beta_\phi[\vis[i]] + U_\phi[\vis[i], \participant[i]]} &&\newcol{\textit{precision submodel}}\\
    \newcol{\logit\left(\pi[i]\right)} &\newcol{= \beta_\pi[\vis[i]] + U_\pi[\vis[i], \participant[i]]} &&\newcol{\textit{zeros submodel}}
\end{align*}
\end{footnotesize}
As with $U[v,p]$ in the previous model, we can similarly use multivariate Normal distributions to model the random offsets:  $U_\mu[v,p]$, $U_\phi[v,p]$ and $U_\pi[v,p]$. It might be tempting to use three separate multivariate Normals for this purpose; however, this would not account for the fact that these offsets are likely correlated across submodels: e.g., someone with a lower mean than average (negative $U_\mu[v,p]$) likely also has a higher probability of getting the answer exactly correct than average (positive $U_\pi[v,p]$). To allow such relationships, we model all random offsets with a shared covariance matrix ($\Sigma$):
\begin{small}

\begin{align*}
    \begin{bmatrix} 
            U_{\newcol{\mu}}[1,p]\\ \vdots\\ U_{\newcol{\mu}}[V,p]\\
            \newcol{U_\phi[1,p]}\\ \newcol{\vdots}\\ \newcol{U_\phi[V,p]}\\
            \newcol{U_\pi[1,p]}\\ \newcol{\vdots}\\ \newcol{U_\pi[V,p]}
        \end{bmatrix}
        &\sim \Normal\left(
            \begin{bmatrix} 
                0\\ \vdots\\ 0\\ 
                \newcol{0}\\ \newcol{\vdots}\\ \newcol{0}\\
                \newcol{0}\\ \newcol{\vdots}\\ \newcol{0}
            \end{bmatrix},
            \Sigma
        \right)
        &&\forall p \in \{1 \dots P\}
        &\begin{array}{l}\textit{correlated}\\ \textit{random offsets}\end{array}
\end{align*}
\end{small}
We build this model in \emph{brms}~\cite{burkner2017brms}, a modeling library in the R statistical programming language~\cite{R}, which fits models using Stan~\cite{carpenter2017stan}, a probabilistic programming language and Markov chain Monte Carlo sampler. The model can be specified using the \texttt{\KeywordTok{brm}} function in modified Wilkinson-Pinheiro-Bates syntax~\cite{wilkinson1973symbolic,pinheiro2013nlme}:
\begin{scriptsize}
\begin{Highlighting}[]
\KeywordTok{brm}\NormalTok{(}
  \KeywordTok{brmsformula}\NormalTok{(}
\NormalTok{    abs_error }\OperatorTok{\codetilde}\StringTok{ }\NormalTok{vis }\OperatorTok{+}\NormalTok{ 0 }\OperatorTok{+}\StringTok{ }\NormalTok{(vis }\OperatorTok{+}\NormalTok{ 0 }\OperatorTok{|}\NormalTok{pt}\OperatorTok{|}\StringTok{ }\NormalTok{participant),}\CommentTok{ # mean}
\NormalTok{          phi }\OperatorTok{\codetilde}\StringTok{ }\NormalTok{vis }\OperatorTok{+}\NormalTok{ 0 }\OperatorTok{+}\StringTok{ }\NormalTok{(vis }\OperatorTok{+}\NormalTok{ 0 }\OperatorTok{|}\NormalTok{pt}\OperatorTok{|}\StringTok{ }\NormalTok{participant),}\CommentTok{ # precision}
\NormalTok{           zi }\OperatorTok{\codetilde}\StringTok{ }\NormalTok{vis }\OperatorTok{+}\NormalTok{ 0 }\OperatorTok{+}\StringTok{ }\NormalTok{(vis }\OperatorTok{+}\NormalTok{ 0 }\OperatorTok{|}\NormalTok{pt}\OperatorTok{|}\StringTok{ }\NormalTok{participant)}\CommentTok{  # zeros}
\NormalTok{  ),}
  \DataTypeTok{family =}\NormalTok{ zero_inflated_beta,}\CommentTok{                        # likelihood}
\NormalTok{  ...}
\NormalTok{)}
\end{Highlighting}
\end{scriptsize}
This syntax closely parallels the equations for the likelihood and the mean ($\mu[i]$), precision ($\phi[i]$), and zeros ($\pi[i]$) submodels. The use of \texttt{\OperatorTok{ +}\NormalTok{ 0}} tells \emph{brms} to use a \emph{one-hot coding} of the \emph{vis} variable: each level of \emph{vis} has its own coefficient. Otherwise, the default \emph{dummy coding} would be used for categorical variables, which would use an intercept for one level of \emph{vis} and coefficients for the offset from that intercept for the other three levels of \emph{vis}. Dummy coding makes it difficult to set the same prior on each visualization type (because one is an intercept and the other three are offsets), whereas \emph{one-hot coding} makes it straightforward (because each is its own intercept). Finally, \texttt{\OperatorTok{|}\NormalTok{pt}\OperatorTok{|}} (where the identifier \texttt{pt} is arbitrary) is used in \emph{brms} in place of a \texttt{\OperatorTok{|}} from standard Wilkinson-Pinheiro-Bates syntax to indicate that the covariance matrix for random effects ($\Sigma$) is shared across submodels.

\subsubsection{Priors}

\begin{figure}
    \centering
    \includegraphics[width=0.47\textwidth]{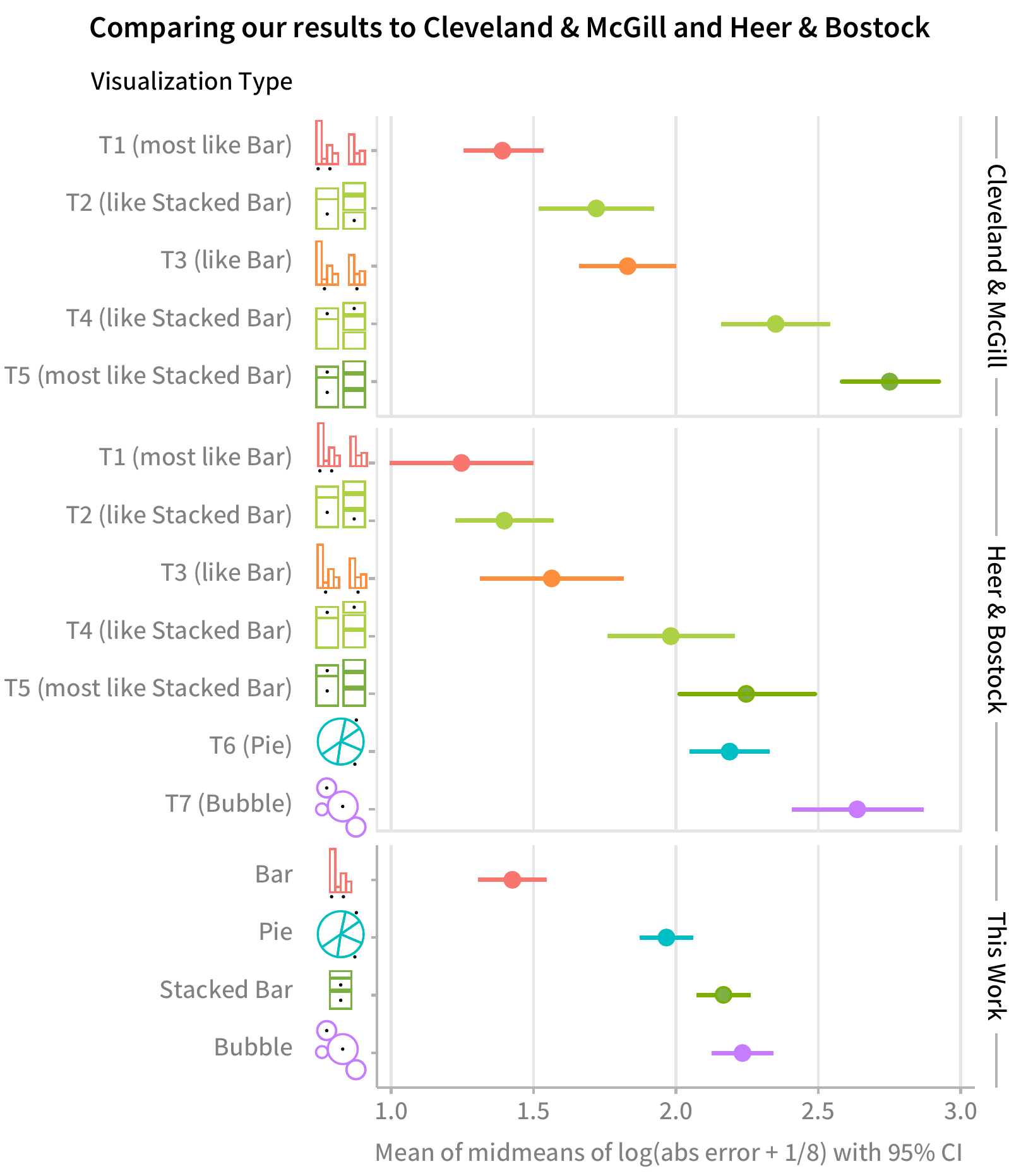}
    \caption{Our results from replicating the analysis in Cleveland \& McGill/Heer \& Bostock; upper two-thirds is a remake of Heer \& Bostock Figure 4.}
    \label{fig:bootrap-replication}
\end{figure}

To fit the Bayesian model we must supply priors for unknown parameters. We used \emph{weakly-informed} priors \cite{gelman2017prior}; that is, priors which are set to cover reasonable ranges of values \emph{a priori}, rather than priors set tightly around a best estimate from previous literature (an \emph{informed} prior). Ideally, this conservative approach allows our priors to regularize \cite{mcelreath2016rethinking} estimates and improve model fit without unduly biasing estimates towards previous results. Our priors are as follows:

\begin{itemize}
\item $\beta_\mu[v] \sim \Normal(-2, 1)$: Prior for the mean error in percentage points on a log-odds scale for each visualization. This prior covers roughly $[-4,0]$ in log-odds space in its 95\% central interval, which is roughly $[1.7,50]$ in percentage points; in other words, we do not expect people to make errors larger than 50 percentage points on average, which would be quite a large error in a proportion judgment task. Prior work also has not found mean absolute error to be credibly less than 1 on the adjusted $\textrm{log}_2$-absolute-error scale of Cleveland \& McGill (Figure \ref{fig:bootrap-replication}); this translates to $2^1 - 1/8 = 1.875$ (inverting their log transformation), which is covered by the 1.7pp lower bound of our prior 95\% interval.
\item $\beta_\phi[v] \sim \Student(5, 0, 10)$: Prior for the precision parameter for each visualization on a log scale. Not having strong prior knowledge of the precision of people's estimates, we chose a wide, relatively heavy-tailed prior (in terms of orders of magnitude, the 95\% central interval of this distribution, back-transformed from the log scale, goes from roughly $1\mathrm{e}{-11}$ to $1\mathrm{e}{11}$).
\item $\beta_\pi[v] \sim \Normal(-2.5, 1.25)$: Prior for the probability of a participant getting a response exactly correct (\ie, a response with 0 error) on a log-odds scale for each visualization. This prior covers roughly $[-5,0]$ in log-odds space in its 95\% central interval, which is roughly $[0.7,50]$ in percentage points. We expected there to be at least about 1\% of zero-error responses, but that it is very unlikely to see more than 50\% of responses being exactly correct.
\item We decompose the covariance matrix of the random effects ($\Sigma$) into standard deviations and a correlation matrix. We set a $\HalfNormal(0,0.5)$ prior on the standard deviations (this is a relatively wide prior as all random effects standard deviations are for coefficients on a log scale), and an $\textrm{Lewandowski-Kurowicka-Joe (LKJ)}(4)$ prior on the correlation matrix~\cite{lkj2009generating}.
\end{itemize}
The fit model had 20 chains with 60,000 total post-warmup samples, all $\hat R \le 1.01$.
The fit was thinned to 6,000 post-warmup samples to expedite calculations, with bulk effective sample sizes of 4,200--6,300 (participant-level variables) and 3,100--5,400 (population-level).

\section{Results}
\label{sec:results}

We begin by replicating analysis methods from prior work on our data as a point of comparison, before using our models to investigate individual differences in performance.

\subsection{Replicating Analyses from Cleveland \& McGill and Heer \& Bostock}\label{result-replicate}

\begin{figure*}[t]
    \centering
    \includegraphics[width=1.0\linewidth]{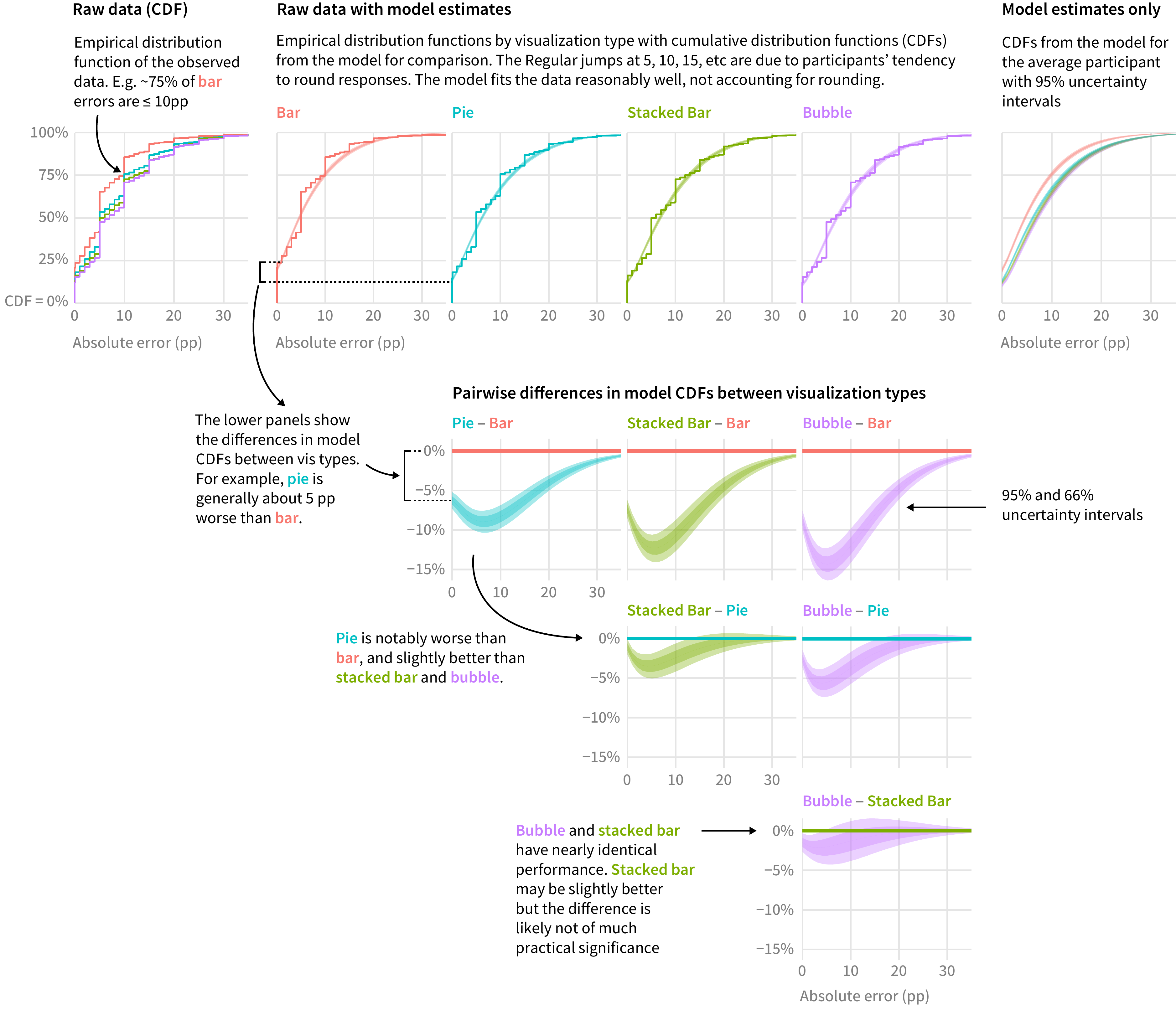}
    \caption{Primary population-level results as \CDFs. Figures are truncated to show errors up to 35pp, which covers 98.5\% of all observed errors.}
    \label{fig:population-summary}
\end{figure*}

To compare our data to previous work, we extracted the reported results from two previous studies: the original Cleveland \& McGill study \cite{cleveland1984graphical} (their Figure 16; upper panel of our Figure \ref{fig:bootrap-replication}), and Heer \& Bostock's crowdsourced experiment \cite{heer2010crowdsourcing} (their Figure 4; middle panel of our Figure \ref{fig:bootrap-replication}). The lower panel of Figure \ref{fig:bootrap-replication} shows an analysis of our data. For the purposes of this comparison, we used methods that match as closely as possible to those of the previous authors: means of midmeans of adjusted log error with (stratified) bootstrapped confidence intervals.

By replicating these analysis methods, we arrive at error estimates and relative visualization rankings that broadly agree with prior work. Visualization type \textit{T3} of prior work is most similar to our \Barv{} type, and all 3 studies arrive at estimates that are essentially identical. Our \Pie{} results match with \textit{T6} of \cite{heer2010crowdsourcing}, and our \Bubble{} results closely match their \textit{T7} type. Finally, our \StackedBar{} type is most similar to \textit{T5} in the two prior studies, and our error estimate is approximately centered between the results estimates from those studies.

Across the three studies, the primary ambiguity is in the exact relative rankings of the visualization types other than \Barv{}. While \Barv{} has the lowest error overall, the remaining rankings are roughly \Pie{} $<$ \StackedBar{} $<$ \Bubble{}. However, all three have qualitatively similar error estimates. As such, it is not clear that one would be justified in holding a strong second-place preference among the three visualization types.

However, this analysis has a number of limitations, particularly when it comes to our goal of comparing participant performance. 
First, due to the use of midmeans, errors made by participants that are far smaller and far larger than average are excluded from consideration, yet the occurrence of such errors may be of great practical concern to visualization designers (particularly large errors). 
Second, the results are reported on an adjusted log scale rather than the original response scale of percentage points, which makes it difficult for a visualization designer to answer questions like, ``how much greater is the expected error between the \Barv{} and \Pie{} chart types, and is that difference practically large enough to influence my design?"
Finally, this analysis methodology produces an average and interval that combines many different people, without providing a direct way of exploring how individuals might vary in performance across visualization types, including whether or not relative rankings differ across people.

\subsection{Distributions: A Step Beyond Averages}

Before considering further analysis, it is important to distinguish between two sources of information available to us: the \textit{observed data}, which are the responses collected directly from individual participants; and the \textit{posterior distribution} of the fitted model, which
encodes
our understanding that people will vary in their responses across repeated trials and conditions, that people vary from each other, that the participants in this experiment are only a sample, and that there may be correlations both between and within people. The model attempts to quantify all of  these sources of uncertainty. 
We show observed data together with model estimates whenever feasible.

With the model and data, we inspect performance based on \textit{error distributions}, rather than average error alone.
The top row of Figure \ref{fig:population-summary} shows the empirical \textit{cumulative distribution functions} (\CDFs) of participant errors as solid, opaque lines. \CDFs{} can provide answers to questions like, \emph{what proportion of the observed errors were less than 5 percentage points (pp)?} \emph{What proportion were less than 10pp?} etc. The shaded bands show the 66\% (dark) and 95\% (light) credible intervals for the \CDF{} of the average participant as estimated by the model.

The \CDFs{} of observed data have several sharp jumps, like stair-steps. This artifact has two causes: 1) of the twenty levels of true proportion examined in this study, eighteen were evenly divisible by 5; 2) roughly 70\% of participant responses were also evenly divisible by 5, i.e. participants tended to round their responses to end in 0 or 5, similar to prior studies (\eg Talbot \etal \cite{talbot2014four}). For example, when the true proportion for a task was 60\%, 373 responses were exactly 55\% (5pp of error), but only 214 responses were between 56\% and 59\%. 
The model does not directly model participant rounding behavior, but the top-center of Figure \ref{fig:population-summary} shows how the model effectively smooths the response curve in a way that still closely fits the data. Since rounding is an artifact of the elicitation procedure, and our model did a good job of otherwise recovering the shape of the distributions, we did not consider it important to model in higher fidelity.

Figure \ref{fig:population-summary} shows that the median error (50\% on the y axis) is around 5pp for all conditions in the observed data.
The distributions are also skewed: the top quartile of errors (75\% on the y-axis) are as much as 2-4 times larger than those in the bottom quartile (25\% on the y-axis). 
Finally, the vast majority of errors are less than 35pp (about 98.5\% below 35pp), regardless of visualization type.

\subsection{Pairwise Comparisons Between Visualization Types}

One issue with the \CDFs{} shown in the top row of Figure \ref{fig:population-summary} is that it is difficult to make comparisons between visualization types to determine their performance relative to each other. 
The bottom half of Figure \ref{fig:population-summary} shows pairwise differences between \CDFs{} of each visualization type. 
For example, where the top row of Figure \ref{fig:population-summary} shows that the average participant is estimated to to make zero-error responses about 25\% of the time with \Barv{}, the pairwise graph labelled \textit{Pie - Bar} shows that the proportion of zero-error responses for \Pie{} will be about 5--7pp less than \Barv{}.
\Barv{} generally dominates \Pie{}, having a higher proportion of smaller errors at every magnitude of error. At error levels higher than 15pp, the difference between \Pie{} and \Barv{} shrinks to less than 5\% of cumulative errors. By an error level of 35pp, the predicted errors for \Pie{} and \Barv{} charts are virtually indistinguishable. 

Overall, Figure \ref{fig:population-summary} shows \Barv{}  has the lowest-error responses for the average participant. \Pie{} is ranked second, receiving proportionally less error than \Bubble{} and \StackedBar, though primarily at error levels below 10pp. There is effectively a tie between \Bubble{} and \StackedBar, with a small probability favoring \StackedBar{} over \Bubble{} at error levels below 20pp; however, the large overlap of zero in the \textit{Bubble - Stacked Bar} graph shows that there can be no certainty that either is better than the other. In any case, the cumulative difference in the two is never estimated to be greater 5\% at any given error level, so there is likely little practical difference between them for the average participant. 
These results largely agree with canonical rankings of these visualization types in prior graphical comparison, while adding perspective about the relative rates of occurrence and magnitudes of errors people make.

\subsection{Between-Person Variance in Mean Error} \label{sec:between-person-variance-in-means}

\begin{figure}[t]
    \centering
    \includegraphics[width=0.47\textwidth]{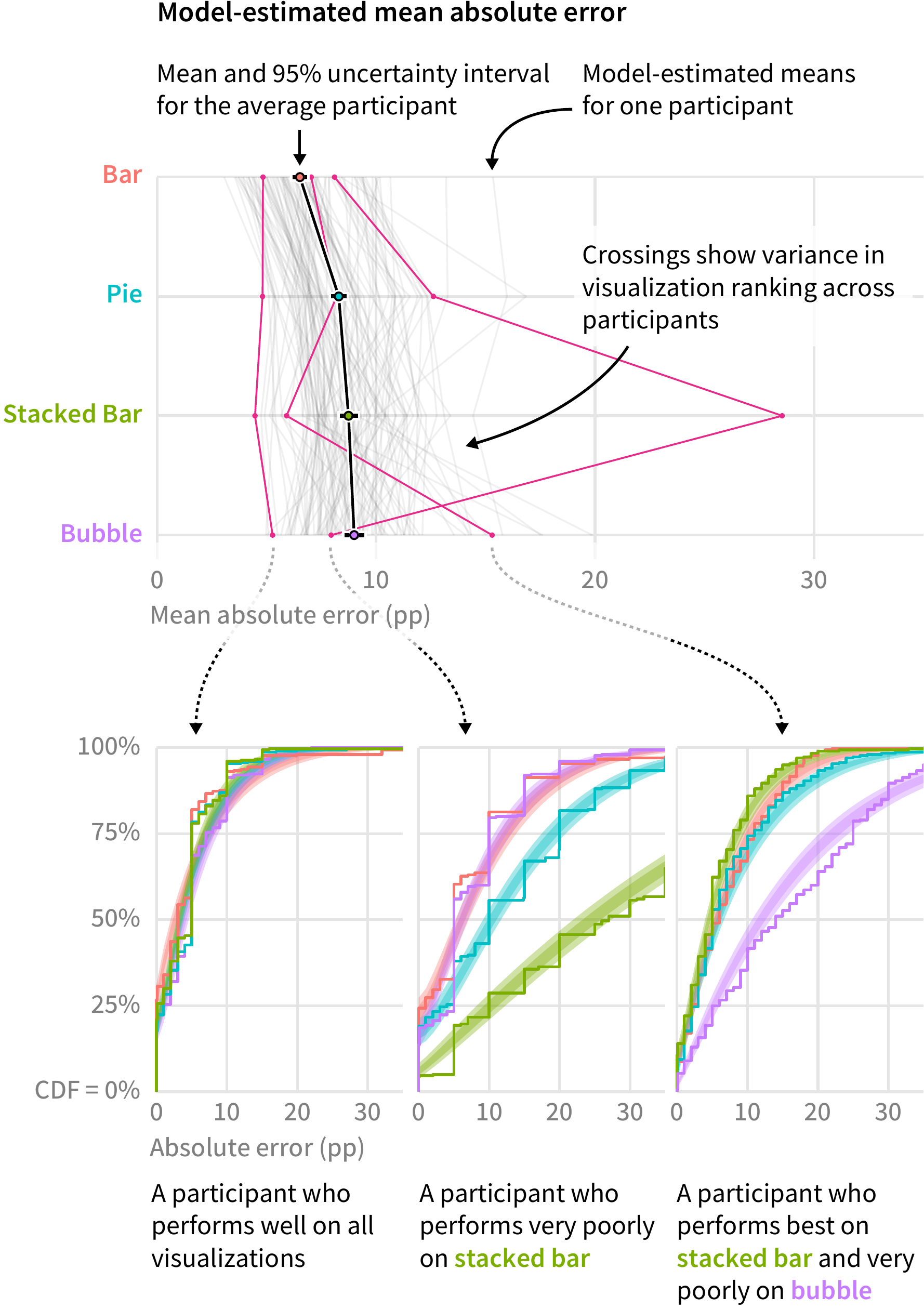}
    \caption{Individual-level mean errors compared to population means as a parallel coordinates chart, with example \CDFs{} from three participants.}
    \label{fig:individual-summary}
\end{figure}

While our results at the population level agree with previous population-level analyses, our model also allows us to examine individual-level performance, asking questions like: \emph{do some people perform at their best on a visualization type that is the worst type of chart for other people? Are the canonical visualization rankings really universal?}

The top of Figure \ref{fig:individual-summary} shows model-estimated means for all participants as gray lines, with the average participant in black. The large between-person variance can be seen as the wide spread of the positions of the gray lines relative to the difference in conditions: in many ways individual differences dwarf the effects of specific chart types. We also highlight three participants with highly ``non-canonical'' rankings: one person who is good at every chart type, one who is particularly bad at \StackedBar, and one who is bad at \Bubble{} and performs best on \StackedBar{} (better even than \Barv). These participants are also not unusual: note the characteristic crossing pattern in the bottom part of the parallel coordinates chart, indicating a large proportion of reversals of the average-participant ranking of \StackedBar{} $<$ \Bubble.

Our hierarchical model allows us to simulate new participants while accounting for the correlation between people's performance across conditions (e.g., that people who are better at one chart type are likely better at others; see \ref{sec:mean-error-correlation}. We can quantify the between-person variance in mean errors by simulating a sample of 6,000 new people within each one of the 6,000 draws from the posterior distribution of our model, then taking the standard deviation of the mean error from each simulated sample of people. This yields 6,000 draws from a distribution describing our uncertainty in the standard deviation of people's mean error.

\begin{figure}[t]
    \centering
    \includegraphics[width=0.47\textwidth]{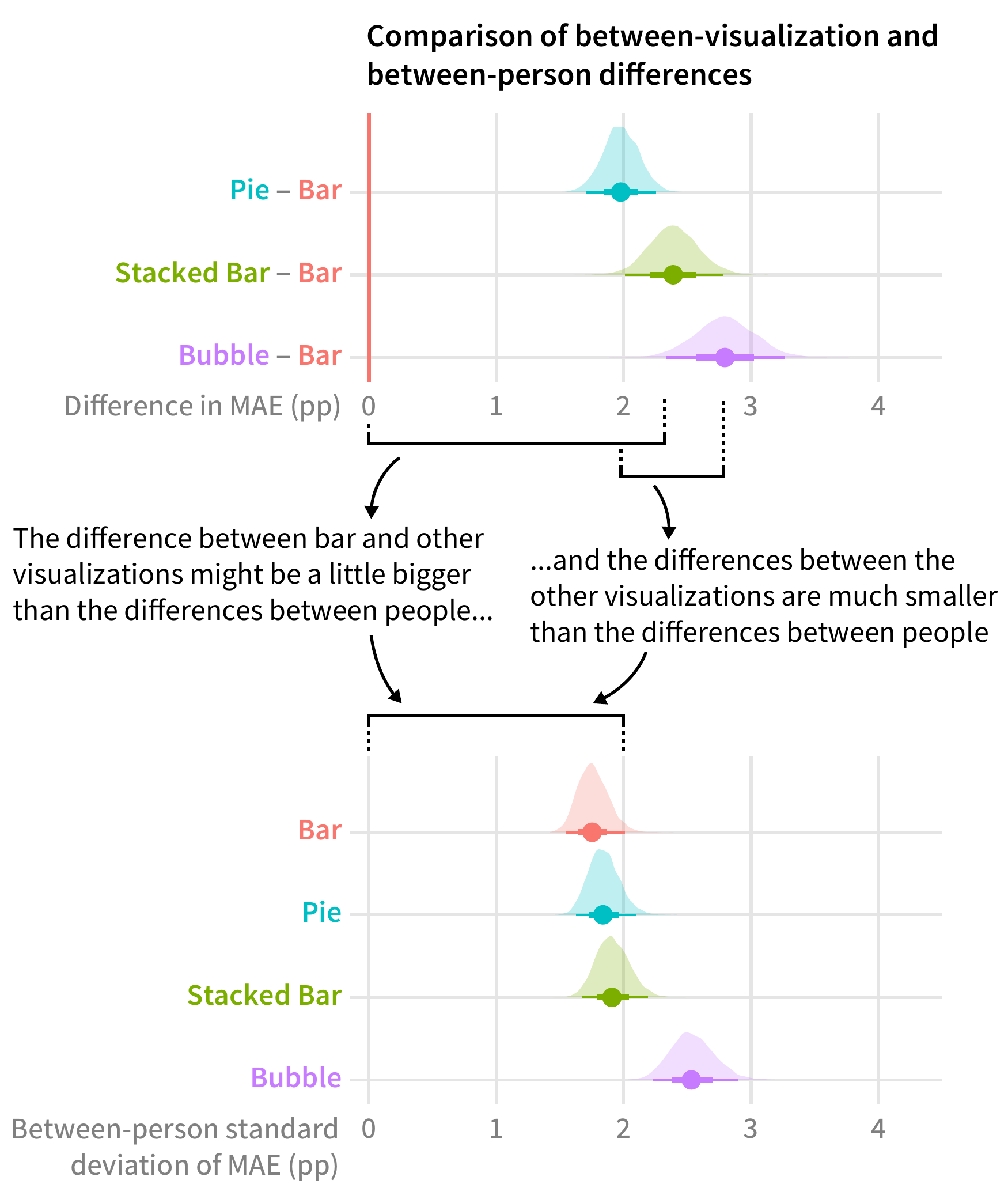}
    \caption{Difference in mean absolute error compared to the \Barv{} condition (top) and standard deviations in individuals' mean absolute errors associated with swapping one participant for another without changing the chart type (bottom), with posterior densities, medians, and 66\% and 95\% uncertainty intervals.}
    \label{fig:variance-between-people-and-charts}
\end{figure}

This provides an estimate of how much variance in error there is expected to be between people.
Figure \ref{fig:variance-between-people-and-charts} compares the differences in mean absolute error between charts types (top) to these standard deviations of between-person mean absolute errors. For example, the difference in mean error between \Barv{} and \Pie{} is about 2pp (top panel); yet, the standard deviation of between-person mean absolute errors is also about 2pp (bottom panel). Thus, the change in error to be expected by randomly selecting a different participant is about the same as the change we would expect by switching from \Barv{} to \Pie{} with an average participant. For visualizations besides \Barv{}, the mean differences are even smaller (less than 1pp; top panel); for these visualization types, it appears that differences between people are larger than differences between conditions.

\subsection{Correlation in Individuals' Mean Error} \label{sec:mean-error-correlation}

\begin{figure}[tb]
    \centering
    \includegraphics[width=0.47\textwidth]{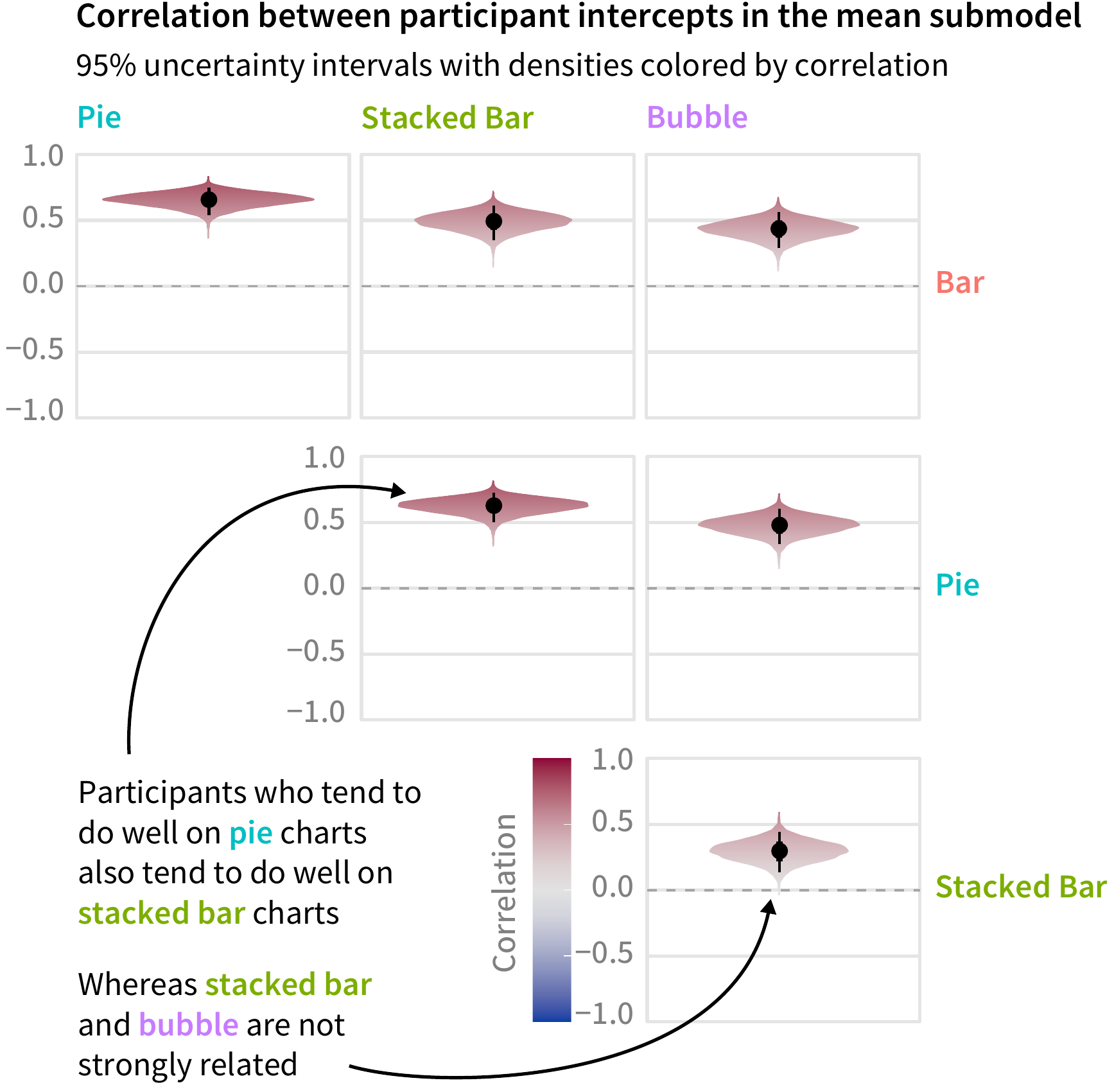}
    \caption{Model-estimated individual-level correlations of visualizations.}
    \label{fig:individual-correlation}
\end{figure}

Examining correlations in the parallel coordinates chart in Figure \ref{fig:individual-summary} suggests another way of looking at the data: correlations between mean error across individuals; e.g., \emph{do people who perform well on \StackedBar{} perform tend to perform well on \Bubble?} As the model estimates correlations between all individual-level parameters in the mean submodel as part of the covariance matrix of hierarchical parameters ($\Sigma$), we can extract these correlations and their uncertainty directly from the model (Figure \ref{fig:individual-correlation}).

Mean error with all visualization types are positively correlated: e.g., if a person does better (worse) than average with \Barv, they will likely do better (worse) than average with \Pie.
The \Barv{} to \Pie{} ($\sim0.65$) and \Pie{} to \StackedBar{} ($\sim0.62$) correlations are highest, while \Bubble{} to \StackedBar{} have much lower correlation. This matches with a qualitative assessment of individual means in Figure \ref{fig:individual-summary}: there is substantial variance from the average ranking of best-to-worst mean error across individuals, particularly in \StackedBar{} and \Bubble.

\subsection{Ranking Individuals' Mean Error} \label{sec:ranking-error}

\begin{figure*}[t]
    \centering
    \includegraphics[width=1.0\linewidth]{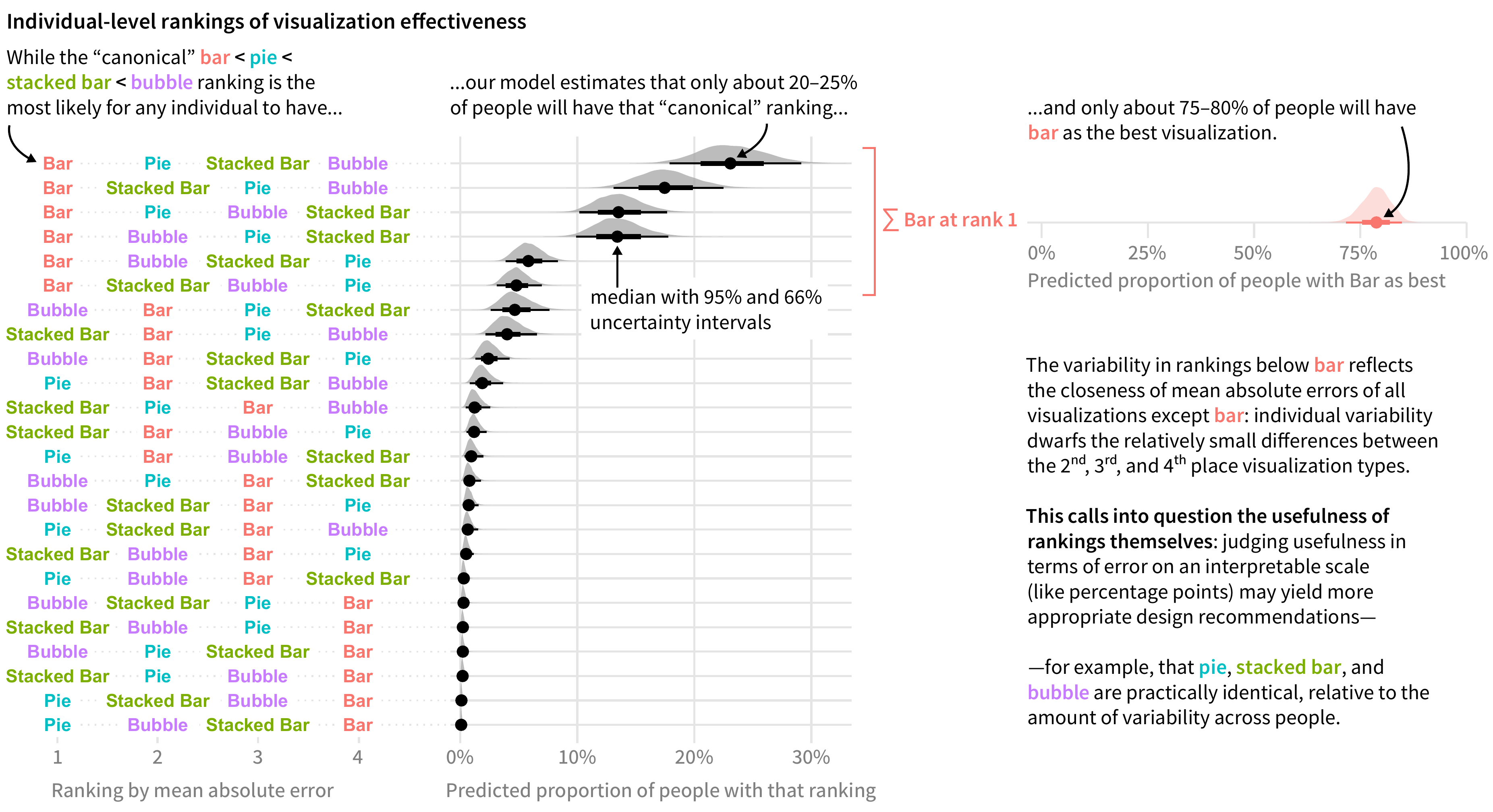}
    \caption{Proportions of the population predicted to have each ranking in terms of mean absolute error, listed in order from most to least common ranking. All proportions added together equal 1 (100\% of the population).}
    \label{fig:ranking-uncertainty}
\end{figure*}

Finally, it is traditional to include an attempt to rank chart types in order to derive actionable design guidance from empirical visualization papers. This task is made difficult by our desire to account for between-person variance; thus, rather than providing a single ranking based on population means, we start by calculating a distribution over rankings. We use the same approach to simulating new samples of people within each draw from our posterior distribution as in 
\ref{sec:between-person-variance-in-means}, but within each draw we calculate the proportion of people having each possible ranking of the four visualization types in terms of mean error. This gives us both the proportion of people expected to have each ranking and the uncertainty in that proportion (Figure \ref{fig:ranking-uncertainty}).

The ``canonical'' ranking of \Barv{} $<$ \Pie{} $<$ \StackedBar{} $<$ \Bubble{} is indeed one shared by the most people---but still only about 22\% of the population. The top-6 most common rankings all have \Barv{} as best, but these account for only about 75--80\% of people. 
There is considerable variance in rankings (particularly for 2nd to 4th place). As our model gives very precise estimates of population-level means, this variance is likely largely driven by large between-person variance as compared to the size of differences between charts types, as we saw in \ref{sec:between-person-variance-in-means}
These results suggest that in the face of large between-person variances, rankings are an inadequate way to summarize visualization performance or derive design guidelines. We discuss alternative approaches to deriving design recommendations in \ref{sec:design-recs}

\section{Discussion}

At face value, graphical perception studies aim to identify the best among a set of visualizations for a given task.
Yet visualization recommendations that build on prior studies are often sparse and conflicting.
One recommendation might suggest that \StackedBar{} charts are a superior choice to \Pie{} charts because our perception of angles is worse than our perception of position.
Another might discourage the use of \Bubble{} charts for similar reasons, citing difficulties in comparing areas.
The results of our replication/extension and modeling efforts suggest that the reality of graphical perception performance is more complex.

\subsection{Between-Person Visualization and Visualization Tasks, Broadly}

One of the primary aims of Cleveland and McGill's study was to investigate whether psychophysical differences predicted differences in chart effectiveness--- which is supported by their aggregate level analysis \cite{cleveland1984graphical}. However, more recent work on more complex chart types has suggested that differences in people's performance cannot simply be attributed to differences in the psychophysical properties in charts: for example, Kale \etal~\cite{kale2020visual} found that the strategies different people use to interpret different uncertainty visualizations had a profound effect on their performance, beyond their ability to estimate the relevant psychophysical quantities (such as ratios of areas in a probability density chart). One could argue that this was an artifact of the complexity of uncertainty visualizations, and that this finding might not replicate for simpler, fundamental tasks like reading ratios in \Barv{} or \Pie{} charts.

Our work suggests this is not the case: even in these fundamental tasks, there is substantial variation in individual performance, and the magnitude of this between-person variance is large compared to the average differences between chart types (see Figure \ref{fig:individual-summary}). Some people are even able to perform equally well (and much better than the population average) on all chart types. Perhaps, as Kale \etal~\cite{kale2020visual} argue for uncertainty visualizations, the particular \emph{strategies} (or \emph{proxies}~\cite{jardine2019perceptual,ondov2020revealing}) people use to accomplish visualization tasks play a larger role in performance, on par with (or in some cases more important than) psychophysical properties of different chart types. 
Studies on pie charts from Kosara \etal underscore this possibility, by establishing that people may be using one of many strategies to perform comparison tasks \cite{kosara2019evidence}.
With the provided modeling approach, such signals for strategy could be investigated as part of future work on an individual, rather than aggregate, basis. 
Furthermore, the provided models could be extended to investigate correlations of individual performance against measures thought to impact visualization performance such as visualization literacy \cite{lee2016vlat, boy2014principled} or spatial ability \cite{ottley2015improving}.

\subsection{Design Recommendations at the Individual Level on the Raw Data Scale} \label{sec:design-recs}

Fitting models to data on the raw error scale which account for individual differences in visualization performance enables us to explore design recommendations for graphical perception across multiple perspectives and scales.

Moving beyond averages to \CDF-based approaches (Figure \ref{fig:population-summary}) allows us to answer questions about possible magnitudes of errors across visualization types and their respective rates of occurrence, along with uncertainty about such estimates. This may enable more nuanced visualization design as it relates to task accuracy.
For example, if the true proportion between two elements of a chart was 50\% (one is half the size of the other), our results suggest that about 98\% of people's responses would be between 15 and 85\%. Whether this magnitude of accuracy could be called ``good enough'' or ``greatly concerning'' would need to be determined by a designer or subject-matter expert on a case-by-case basis.

For example, if a designer is interested in ``all of the errors people are likely to make'', then we might want to look at something like the 75th percentile (or higher) of the distribution. 
Or a designer might say, ``I want to maximize the likelihood of highly accurate judgments'', in which case we could compare the 25th percentile of each distribution, as this would tell us which visualization is most likely to elicit errors that are very small.
Or a designer might say, ``I just want to pick the visualization that is most likely to result in the lowest error judgments most of the time'', which puts us back to comparing medians (50th percentile). 
Such considerations are all but impossible if only summarized averages are made available.

As we saw in \ref{sec:ranking-error}, considering between-person variance also calls into question the value of ``ranking'' visualizations by effectiveness. Looking at between-person variance of errors on the raw scale offers an effective alternative: results suggest that for most, \Barv{} will be about 2 percentage points better than the other three visualization types, and that the remaining differences between visualization types are likely washed out by between-person variance. Thus, a simple recommendation to designers might be: \textbf{use \Barv{} if an extra 2pp of precision is needed; otherwise, the chart type is unlikely to make much difference when factoring in the larger differences between people}.

Analysing error on the raw scale thus makes it easier for designers to incorporate results from the literature into a design process where ``best encoding'' has to compete with other concerns (aesthetics, use of metaphor or rhetoric~\cite{hullman2011rhetoric}, memorability~\cite{borkin2013makes}, etc). It is very hard to make an informed design decision if one just has a ranking of effectiveness: as a designer, one needs to know the magnitude of differences in effectiveness on an understandable scale (ideally the data scale) to make these tradeoffs carefully, and log error abstracts away this understandability. Log error was adopted by Cleveland and McGill as a data analysis convenience, not because it particularly aids interpretation or generalization; \textbf{we suggest the field abandon log error in favor of measures more easily translated into practice}.

\subsection{Between-Person Variance as a form of Visualization Literacy}

Looking at distributions of individual-level mean error, as in Figure \ref{fig:individual-summary}, could allow designers to make judgments about variance between people. How many people are likely to be ``left behind'' by a particular encoding choice? Further research is vital to fully understanding the variance of individuals' performance across the range of visualization types that exist in the literature, in order to understand just how broadly accessible a visualization is. Work on visualization literacy has begun to address this problem~\cite{lee2016vlat,borner2019data,galesic2011graph,boy2014principled,alper2017visualization,chevalier@2018literacy, firat2022interactive}; our work suggests some effects of individual differences may even dwarf effects of different chart types.

The ability to quantify between-person variance in visualization performance raises new possibilities for ongoing efforts in visualization literacy.
For example, chart interpretation strategy may be one explanation for the observed credible differences in participant performance, such as people who perform poorly with the \StackedBar{} but well with all other chart types (\eg Figure \ref{fig:individual-summary}).
Another potential literacy-focused application of the models described here would be to use them to drive feedback and educational interventions, to make people aware of opportunities to improve their skill and reliability in interpreting visualizations.
Efforts in improving ``low-level'' visualization literacy might leverage prior work exploring visualization interpretation strategies, such as arcs, angles, areas for \Pie{} charts \cite{kosara2019evidence}, or perceptual proxies for \Barv{} charts \cite{jardine2019perceptual,ondov2020revealing}.

Beyond improving design recommendations, combined models of variance in individual performance and visualization literacy might also be used to drive \emph{adaptive user interfaces} (user interfaces which adapt to individual characteristics). 
Adaptive user interfaces typically employ models targeting specific personal attributes, taking into account a range of ways in which people may differ---whether it be in abilities (both physical and perceptual)~\cite{gajos2010automatically, peissner2012myui}, preferences as influenced by culture~\cite{reinecke2011mocca}, the users' surroundings and personal characteristics~\cite{peissner2012myui, gasparini2010combining, gajos2005preference}, or their demographic profile~\cite{reinecke2014quantifying}. 
In the visualization community, the Draco project from Moritz \etal could plausibly be extended to take as input \emph{probabilistic} instead of deterministic rules, opening up new avenues for visualization recommendation that accommodate model-based individual differences \cite{moritz2018formalizing}.
Probabilistic representations of rank data, potentially drawing on techniques such as hypothetical outcome plots (HOPs) \cite{kale2018hypothetical}, might also be explored as a means for presenting ranks of visualization performance to designers while faithfully representing individual variance like those modeled here.
Future work might aim to better understand individual traits and their relationship with individuals' performance, developing visualization systems that better optimize accessibility for wider audiences.

\section{Limitations}
While the resulting model reflects error patterns we observed on a set of common visualizations, it is not without limitations, raising questions of validity and reliability as we move beyond the particular visualizations and task considered.
For example, a new visualization type with the same task could lead to markedly different patterns of errors, which could require changes in the model to accommodate.
We ultimately used a Bayesian zero-inflated beta model, which in prior simulation studies has been shown to be more reliable than other models on similar types of responses~\cite{liu2018review}. We also compared this response distribution to other reasonable alternatives (e.g., hurdle lognormal models) and found our substantive conclusions were not sensitive to the response distribution. That said, it is possible some unmodelled (or mis-specified) aspects of people's behavior could have an impact on error in ways not accounted for by our model.
At the same time, further refining resulting models might be considered a strength of model-based analysis, as it would suggest differences in chart reading require different statistical constructs to be adequately captured.

Another limitation is that model-based methods may require additional effort, in part due to its unfamiliarity in research communities that have established norms, practices, and expectations surrounding analyses and reporting.
One of the goals of the present paper is to explore these differences by replicating a familiar visualization study, and to demonstrate some of the possible benefits of model-based approaches for answering research questions that would be difficult or impossible to address through traditional methods.

\section{Conclusion}

In this paper, we undertook the ``substantial chore''~\cite{cleveland1984graphical} of modeling variance and correlation in individuals' performance on classical graphical perception tasks. Our work identifies problems with two common practices in visualization research: (1) modeling or reporting visualization rankings only for the ``average observer'' and (2) reporting only log error. The problem with the first practice is revealed by our finding that substantial between-individual variance exists for even these elementary visualization tasks; \eg, as much as 30\% of people are likely not ``best with the \Barv'', and different people may depart substantially from the canonical ranking of visualization type effectiveness. The problem with the second practice comes from the increasing consideration of factors other than raw effectiveness---aesthetics, metaphor, etc.---in visualization design. 
We now have the tools to analyse errors on the raw scale, laying groundwork for future studies evaluating how visualization designers interpret effect sizes on different scales (\eg log versus original units), and how designers reason about the cost/benefits of optimizing for one measure over another.
Ultimately, we believe the field should move beyond the use of rankings prevalent in prior work, building a more complete picture of the spectrum of human performance on visualization tasks so that we can create more practically-applicable recommendations for visualization designers, and support the important work of measuring and promoting visualization literacy.

\ifCLASSOPTIONcompsoc
  \section*{Acknowledgments}
\else
  \section*{Acknowledgment}
\fi

This work was supported in part by
a grant from the US National Science Foundation (\#1815587, \#1815790).

\ifCLASSOPTIONcaptionsoff
  \newpage
\fi

\bibliographystyle{IEEEtran}
\bibliography{main.bib}

\newpage
\begin{IEEEbiography}
[{\includegraphics[width=1in,height=1.25in,clip,keepaspectratio]{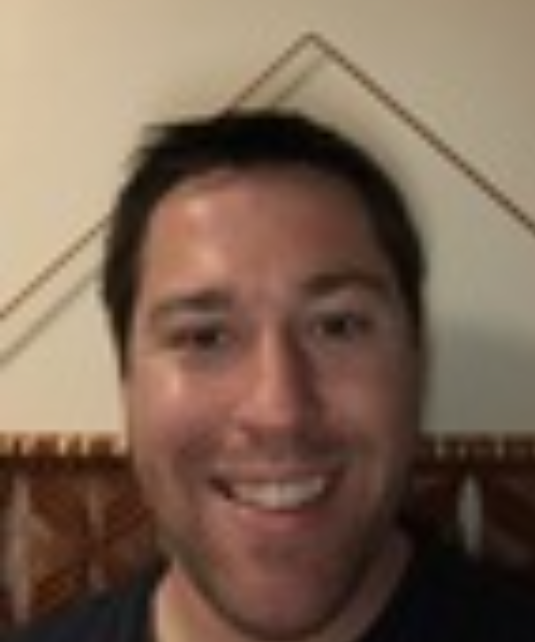}}]
{Russell Davis}  is a Master of Science student majoring in Data Science at Worcester Polytechnic Institute. He obtained his Bachelor’s degree in Chemistry from Clarkson University. His research interests include visualization literacy and visualization for data science applications. Before joining WPI, Russell served in the United States Marine Corps.
\end{IEEEbiography}

\vskip -2\baselineskip plus -1fil

\begin{IEEEbiography}
[{\includegraphics[width=1in,height=1.25in,clip,keepaspectratio]{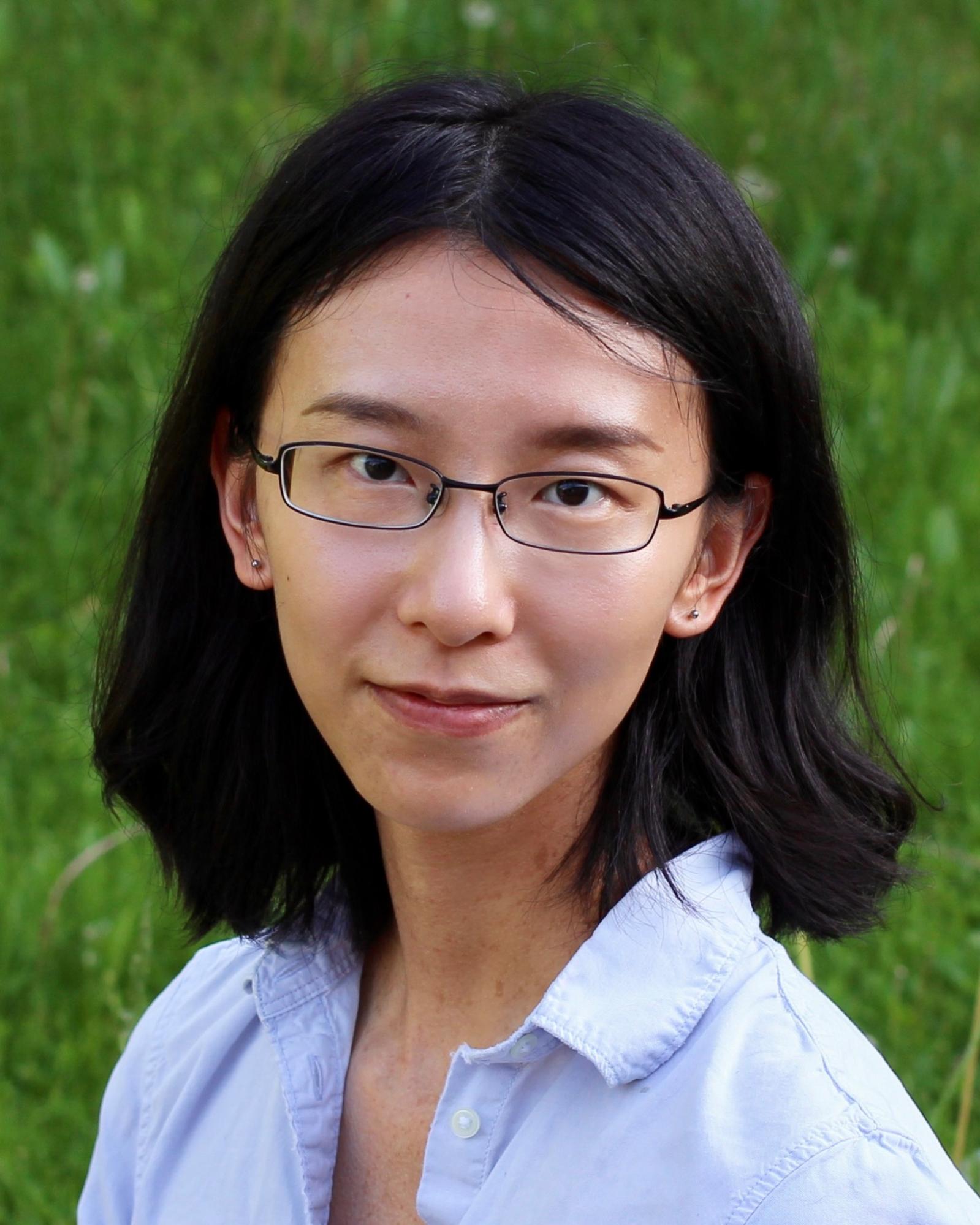}}]
{Xiaoying Pu} is a postdoctoral researcher at the University of California, Merced. She obtained her Ph.D. in Computer Science and Engineering from the University of Michigan, Ann Arbor. Her research interests include uncertainty visualization, visualization grammar, and visual analytics. Her dissertation focused on understanding and building visualization tools for data analysts.
\end{IEEEbiography}

\vskip -2\baselineskip plus -1fil

\begin{IEEEbiography}
[{\includegraphics[width=1in,height=1.25in,clip,keepaspectratio]{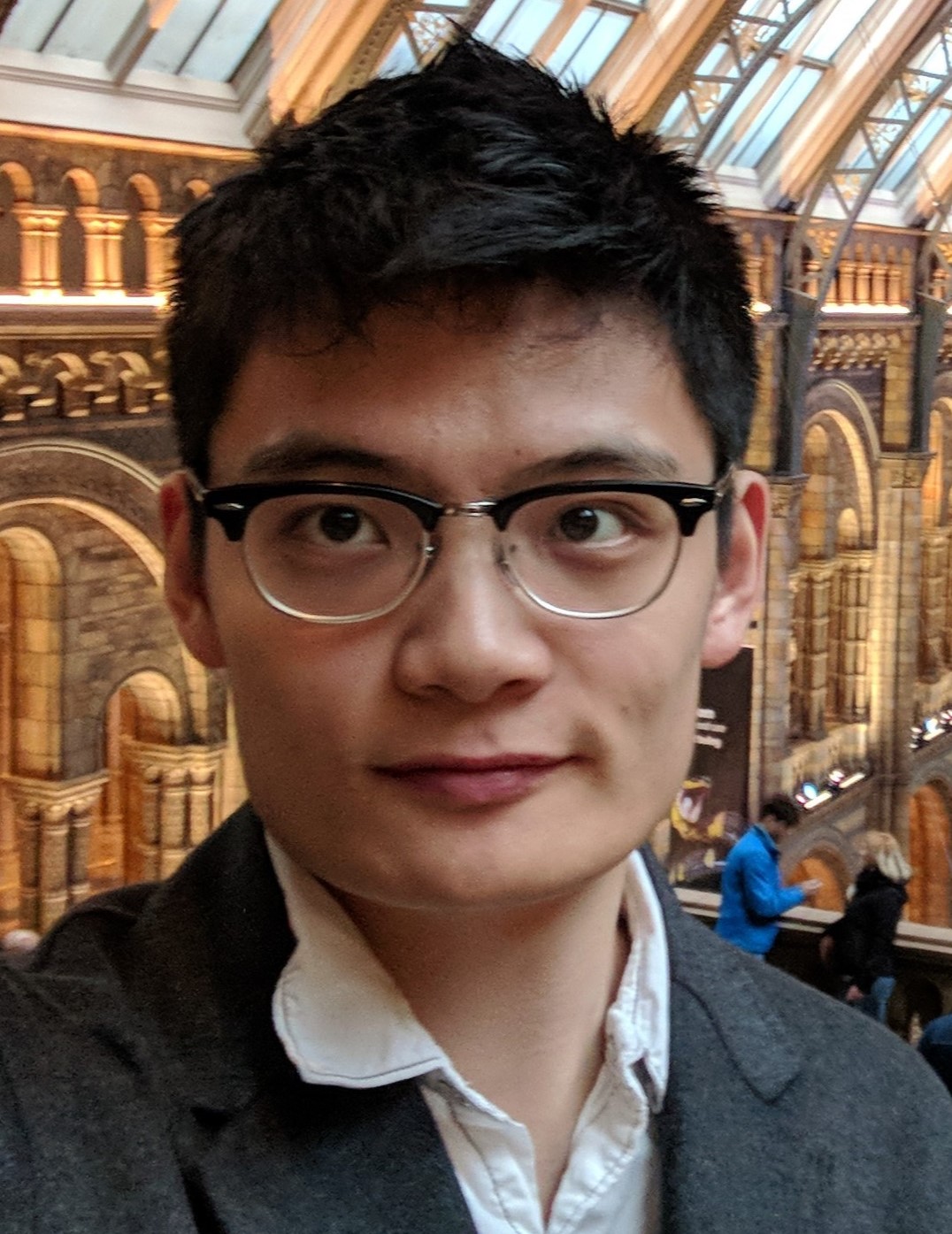}}]
{Yiren Ding} is a Ph.D Student in Computer Science at Worcester Polytechnic Institute (WPI) and a research assistant at the VIEW group. His research focuses on data visualization literacy, animation, and building interactive data visualizations. Before joining WPI, he obtained his Master's degree in Computer Science at Northeastern University and Bachelor's degree from the China University of Geosciences.
\end{IEEEbiography}

\vskip -2\baselineskip plus -1fil

\begin{IEEEbiography}
[{\includegraphics[width=1in,height=1.25in,clip,keepaspectratio]{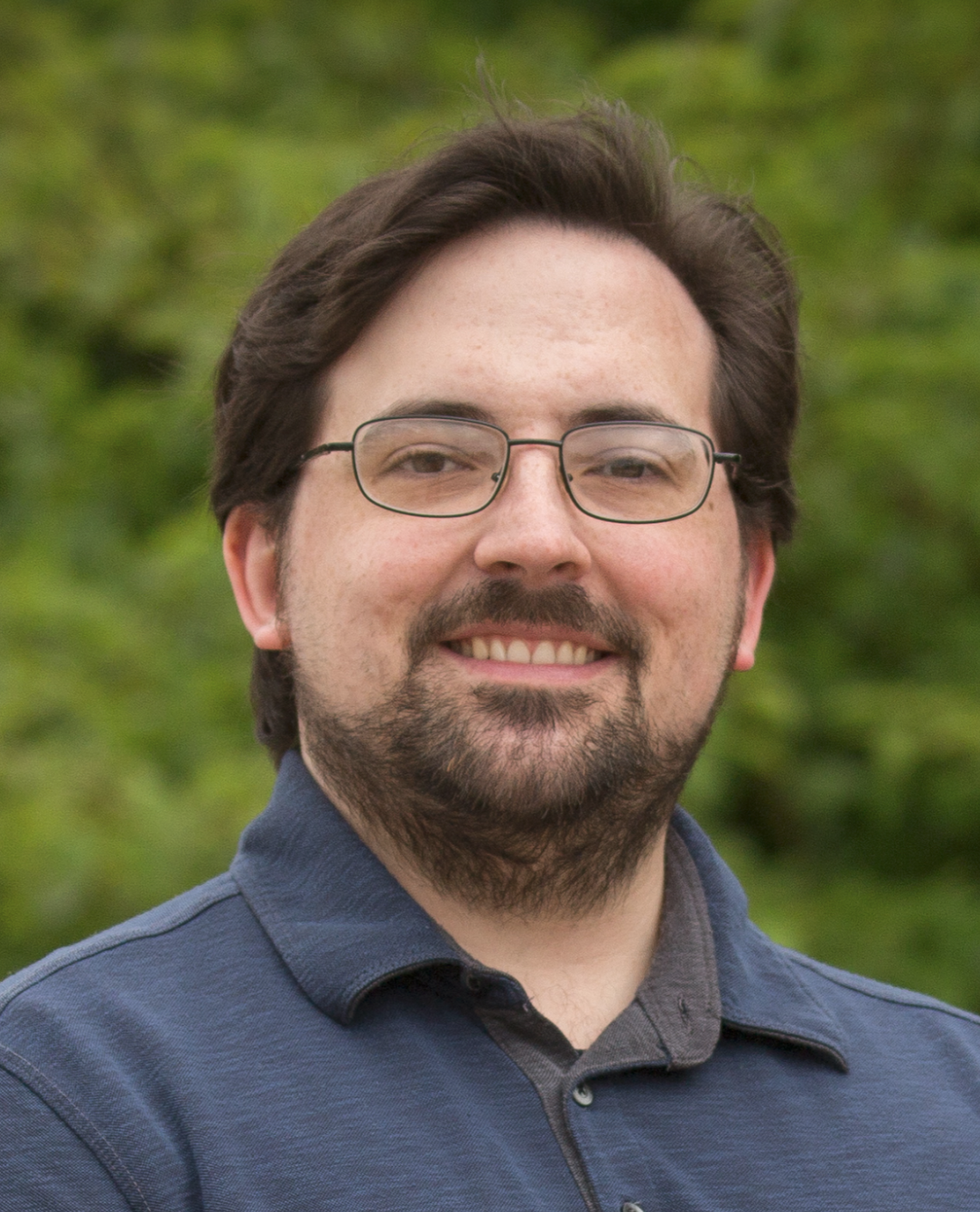}}]
{Brian D. Hall} is a Ph.D Candidate in Information at the University of Michigan, and a Graduate Research Fellow of the National Science Foundation. His work in Human Computer Interaction explores ways to analyze and communicate uncertainty through data visualization and interactive system design. He obtained his Bachelor’s degree in Computer Information Systems and Psychology from the University of Wisconsin - Stevens Point.
\end{IEEEbiography}

\vskip -2\baselineskip plus -1fil

\begin{IEEEbiography}
[{\includegraphics[width=1in,height=1.25in,clip,keepaspectratio]{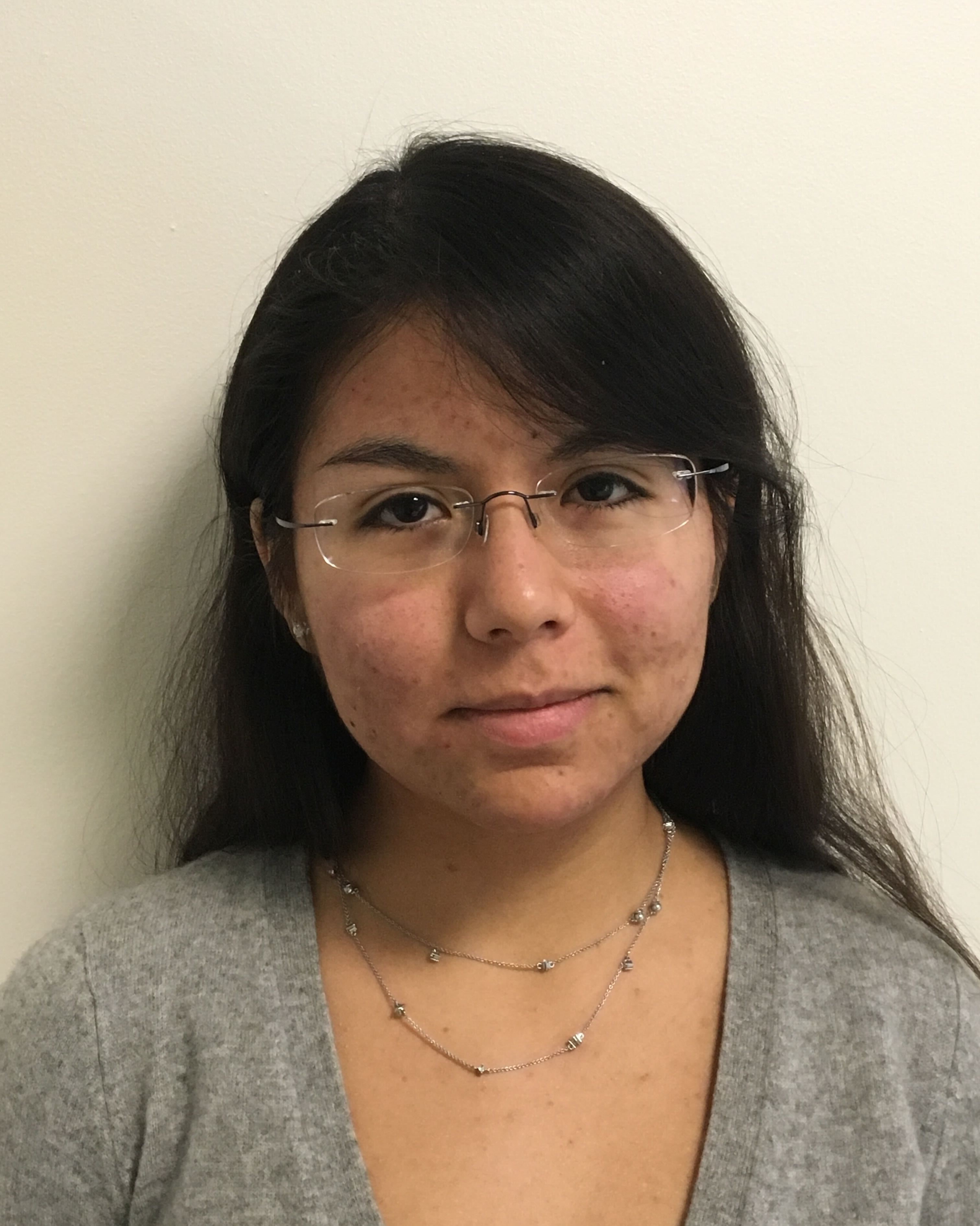}}]
{Karen Bonilla} is a Postgraduate Researcher with the VIEW group in the Department of Computer Science at Worcester Polytechnic Institute. Her work focuses on improvements in teaching visualization literacy as part of middle school curriculums, and building on the current use of visualizations in teaching middle school math. She obtained her Bachelor of Science in Business Administration with a concentration in Economics from Babson College.
\end{IEEEbiography}

\vskip -2\baselineskip plus -1fil

\begin{IEEEbiography}
[{\includegraphics[width=1in,height=1.25in,clip,keepaspectratio]{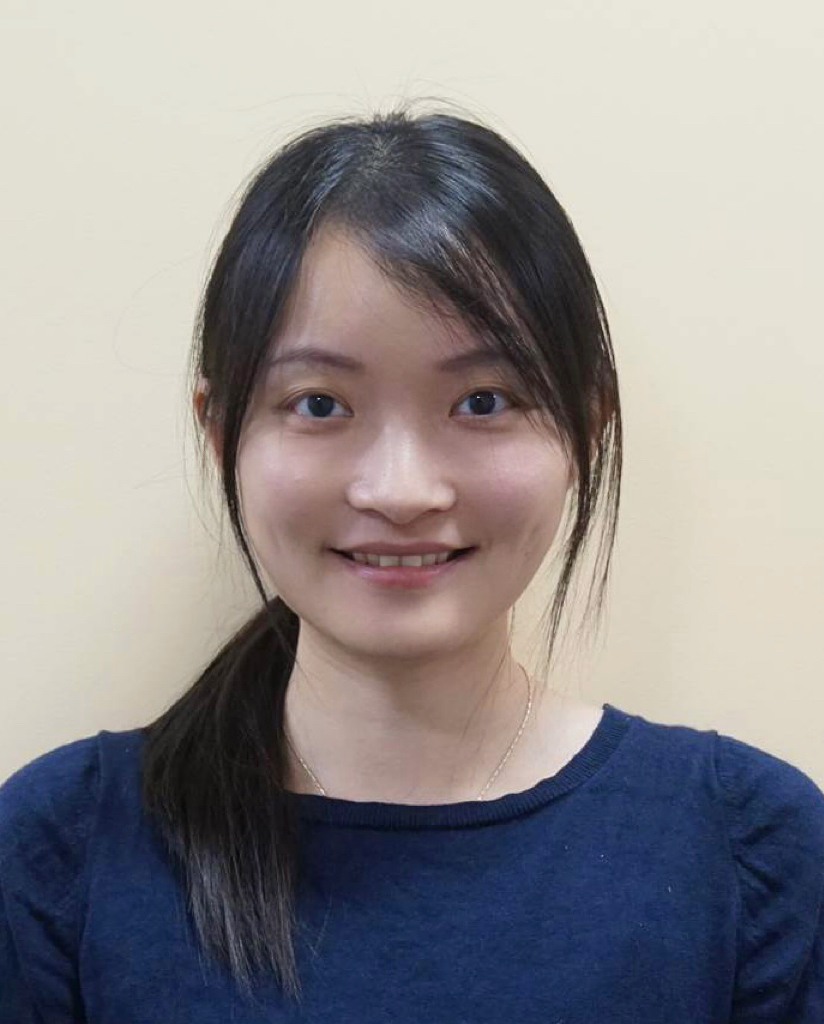}}]
{Mi Feng} obtained her Ph.D degree in computer science from Worcester Polytechnic Institute. Her dissertation focused on understanding and supporting how people interact with data visualizations on the web. Her research interests include information visualization, visual analytics and human-computer interaction.
\end{IEEEbiography}
\vskip -2\baselineskip plus -1fil

\begin{IEEEbiography}
[{\includegraphics[width=1in,height=1.25in,clip,keepaspectratio]{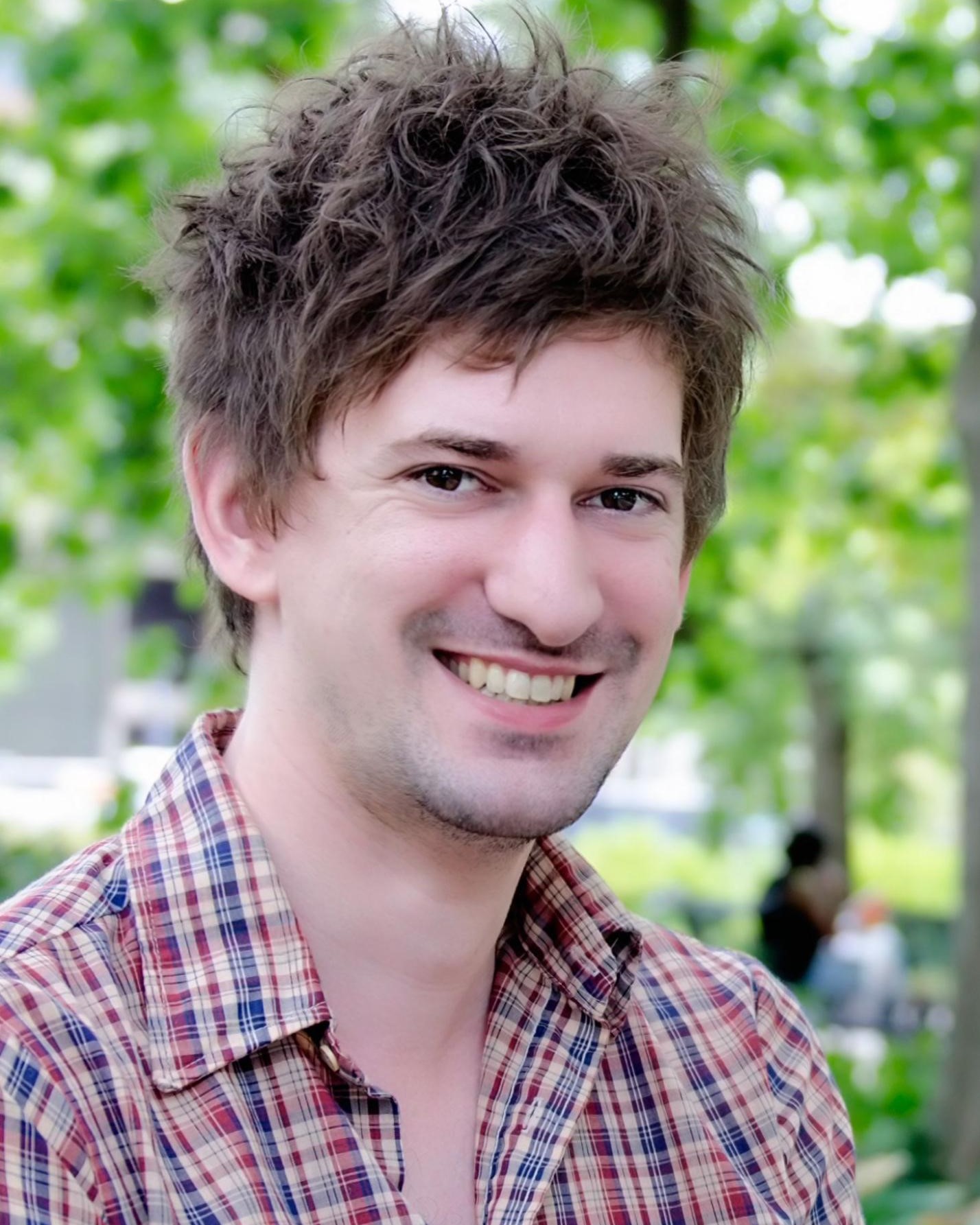}}]
{ Matthew Kay} is an Assistant Professor jointly appointed in Computer Science and Communications Studies at Northwestern University. He works in human-computer interaction and information visualization, and particularly in uncertainty visualization, personal health informatics, and the design of human-centered tools for data analysis. He co-directs the Midwest Uncertainty Collective (http://mucollective.co) and is the author of the tidybayes (https://mjskay.github.io/tidybayes/) and ggdist (https://mjskay.github.io/ggdist/) R packages for visualizing Bayesian statistical model output and uncertainty.
\end{IEEEbiography}
\vskip -2\baselineskip plus -1fil

\begin{IEEEbiography}
[{\includegraphics[width=1in,height=1.25in,clip,keepaspectratio]{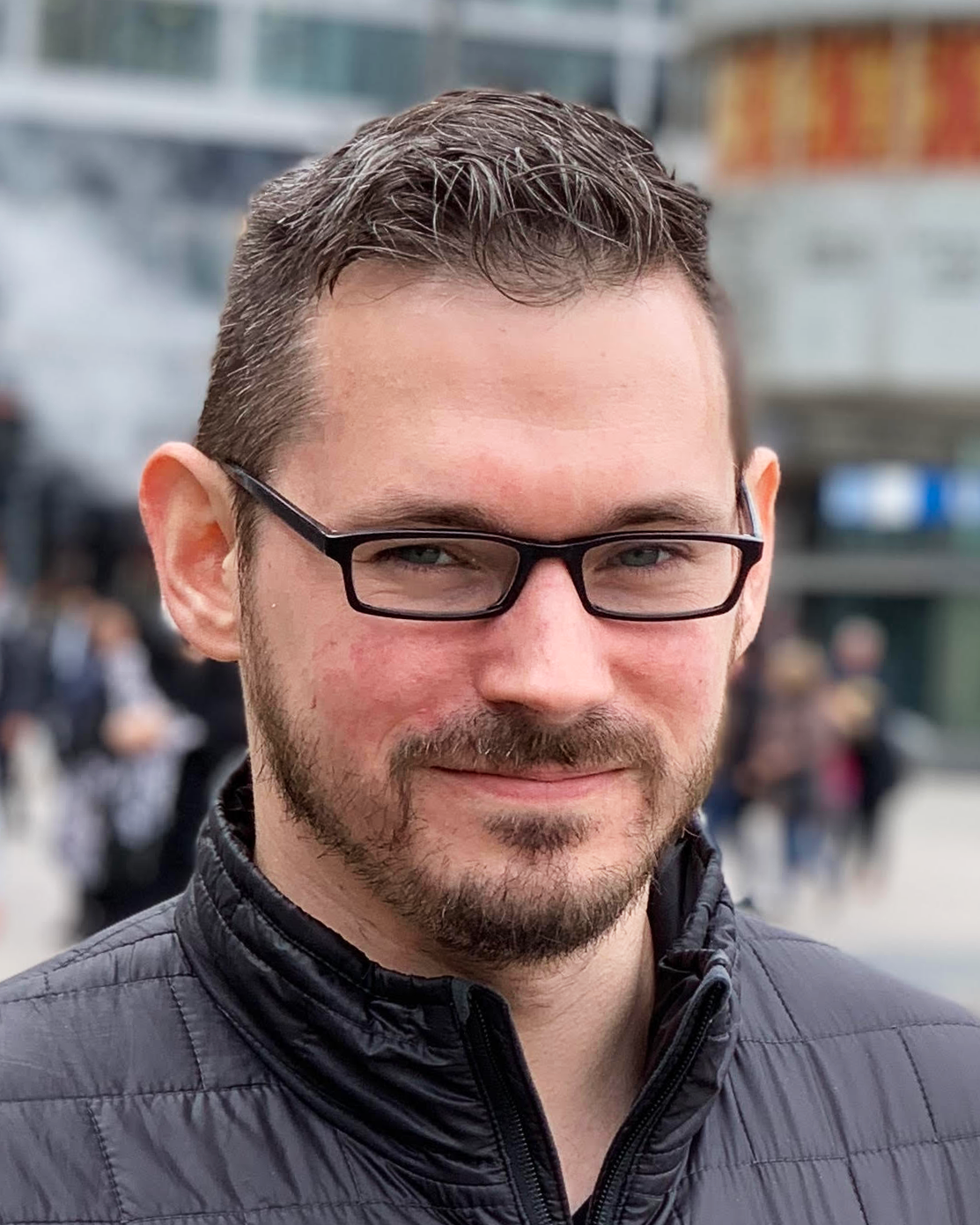}}]
{Lane Harrison} is an Associate Professor in the Department of Computer Science at Worcester Polytechnic Institute. Prior to joining WPI, he was a postdoctoral fellow in the Department of Computer Science at Tufts University. He obtained his Bachelor’s and PhD degrees in computer science from the University of North Carolina at Charlotte. Lane directs the VIEW group at WPI, where he and his students leverage computational methods to understand and shape how people use visualizations and visual analytics tools.
\end{IEEEbiography}

\end{document}